\documentclass[prd,aps,showpacs,nofootinbib,superscriptaddress,preprintnumbers,epsf,psf]{revtex4}

\usepackage[utf8]{inputenc}
\usepackage[T1]{fontenc}
\usepackage{slashed}
\usepackage{amssymb,amsfonts,amsmath,amsthm}
\usepackage{dsfont,bbm}
\usepackage{dcolumn}
\usepackage{epsf}
\usepackage{graphicx}
\usepackage[caption=false]{subfig}
\usepackage{dsfont}
\usepackage{slashed}
\usepackage[active]{srcltx}
\usepackage[usenames]{color}

\begin{document}

\title{New contributions to gluon poles in
direct photon production}

\author{I.~V.~Anikin}
\email{anikin@theor.jinr.ru}
\affiliation{Bogoliubov Laboratory of Theoretical Physics, JINR,
             141980 Dubna, Russia}
\author{O.~V.~Teryaev}
\email{teryaev@theor.jinr.ru}
\affiliation{Bogoliubov Laboratory of Theoretical Physics, JINR,
            141980 Dubna, Russia}

\begin{abstract}
We consider the direct photon production in two hadron collision with  one of the hadrons
being transversely polarized.
By using the contour gauge for gluon fields, we find that there are
new twist-$3$ terms present in the hadron tensor of the considering process in
addition to the standard twist-$3$ terms.
In this work, we demonstrate that the significance of these new terms are
two-fold: first, they are crucial to both the QED and QCD gauge invariance and, second,
their contributions to the hadron tensor are at least the same as those from the standard ones.
We also study the resuting effects which are responsible for the universality breaking of the
corresponding twist-$3$ parton distributions.
\end{abstract}
\date{\today}
\maketitle

\section{Introduction}
\label{Introduction}

The problem of the electromagnetic (QED) gauge invariance in the
deeply virtual Compton scattering (DVCS) and similar exclusive processes
has intensively been discussed during the last few years (for example, see
\cite{Gui98, Pire, APT-GI, Bel-Mul, PPSS}).
This development explored the similarity with the earlier studied inclusive spin-dependent
processes \cite{Efr}.
The gauge invariance of hard process amplitudes is
ensured by twist three contributions and by the use of the
equations of motion that provide a possibility to
exclude the three-particle (quark-gluon) correlators from the amplitude.
So that, after combining with the two-particle
correlator contributions, one gets the
gauge invariant expressions for the physical amplitudes or, in the case of lepton-hadron processes,
for the corresponding hadron
tensors \cite{Efr}. This scheme was originally developed in the case of the particular inclusive processes
with transverse polarized hadrons, like structure function $g_2$ in
DIS \cite{Efr} and Single Spin Asymmetry (SSA)
 \cite{EKT} due to the soft quark
(fermionic poles \cite{Efremov:1984ip}).
{Also, the QCD gauge invariance of the so-called gluonic poles contributions \cite{b}
has been a subject of studies in \cite{Boer:2003cm} where the methods that are used rely on the Wilson exponentials
\cite{Collins:2002kn,Belitsky:2002sm,Collins:2004nx,Cherednikov:2008ua}.

We have shown in our recent work \cite{AT-10} that, in order to ensure the QED gauge invariance of
the transverse polarized Drell-Yan (DY) hadron tensor,
it is mandatory to include
the contribution from the extra diagram originating from the non-trivial
imaginary part of the corresponding twist-$3$ function $B^V(x_1,x_2)$.
Before this study \cite{AT-10}, the function $B^V(x_1,x_2)$ has been
argued as real, and
the imaginary part of amplitude was ensured by means of a specially introduced ``propagator''
\footnote{This is the so-called special propagator originally
suggested by J.w.~Qiu.}
in the hard part of the hadron tensor \cite{BQ}.
However, we have explained in \cite{AT-10} that the
$B^V$-function does, in fact, have an imaginary part, and the existence of this imaginary part
can be realized with a help of the contour gauge.
Moreover, the $B^V$-function with the complex prescription
induces a new contribution to the hadron tensor.
As has been stressed, this extra contribution
leads to the amplification of corresponding tensor by a factor of $2$.
This finding of ours has independently been confirmed in \cite{Ratcliffe13} by
using of a different approach.}
Besides, from the point of view of phenomenology, the corresponding SSAs in the DY
process and the role of gluon pole contributions have previously been
discussed in \cite{P-R, Carlitz:1992fv, Brandenburg:1995pk, Bakulev:2007ej, Radyushkin:2009zg,
Polyakov:2009je, Mikhailov:2009sa, Brandenburg:1994wf, Teryaev, Boer, Ter00,
Ratcliffe:2009pp, Cao:2009rq, Zhou:2009jm, Ma:2003ut}.

In the present paper, we extend our approach used in \cite{AT-10} to the case of the
direct photon production in two hadron collision where one hadron is  transversely polarized.
We derive the hadron tensor for this process and study
the effects which lead to the soft breaking of factorization
(or the universality breaking) through
the QED and QCD gauge invariance.
In a similar manner as in \cite{AT-10}, the special role is played by the contour gauge for gluon fields.
We demonstrate that the prescriptions for the gluonic poles in the twist $3$ correlators
are dictated by the prescriptions in the corresponding hard parts.
We argue that the prescriptions in the gluonic pole contributions differ from each other
depending upon the initial or final state interactions of the diagrams under consideration.
Moreover, the different prescriptions  are needed to ensure the QCD gauge invariance.
We treat this situation as a breaking of the
universality condition resulting in factorization soft breaking.
The extra diagram contributions, which naively do not have an imaginary phase, is discussed in detail.

In the paper, we also show that the new (``non-standard") terms
do contribute to the hadron tensor exactly as the ``standard" terms known previously.
This is exactly similar to the case of Drell-Yan process studied in \cite{AT-10}.

\section{Getting started: case of Drell-Yan process}

\begin{figure}[t]
\centerline{\includegraphics[width=0.3\textwidth]{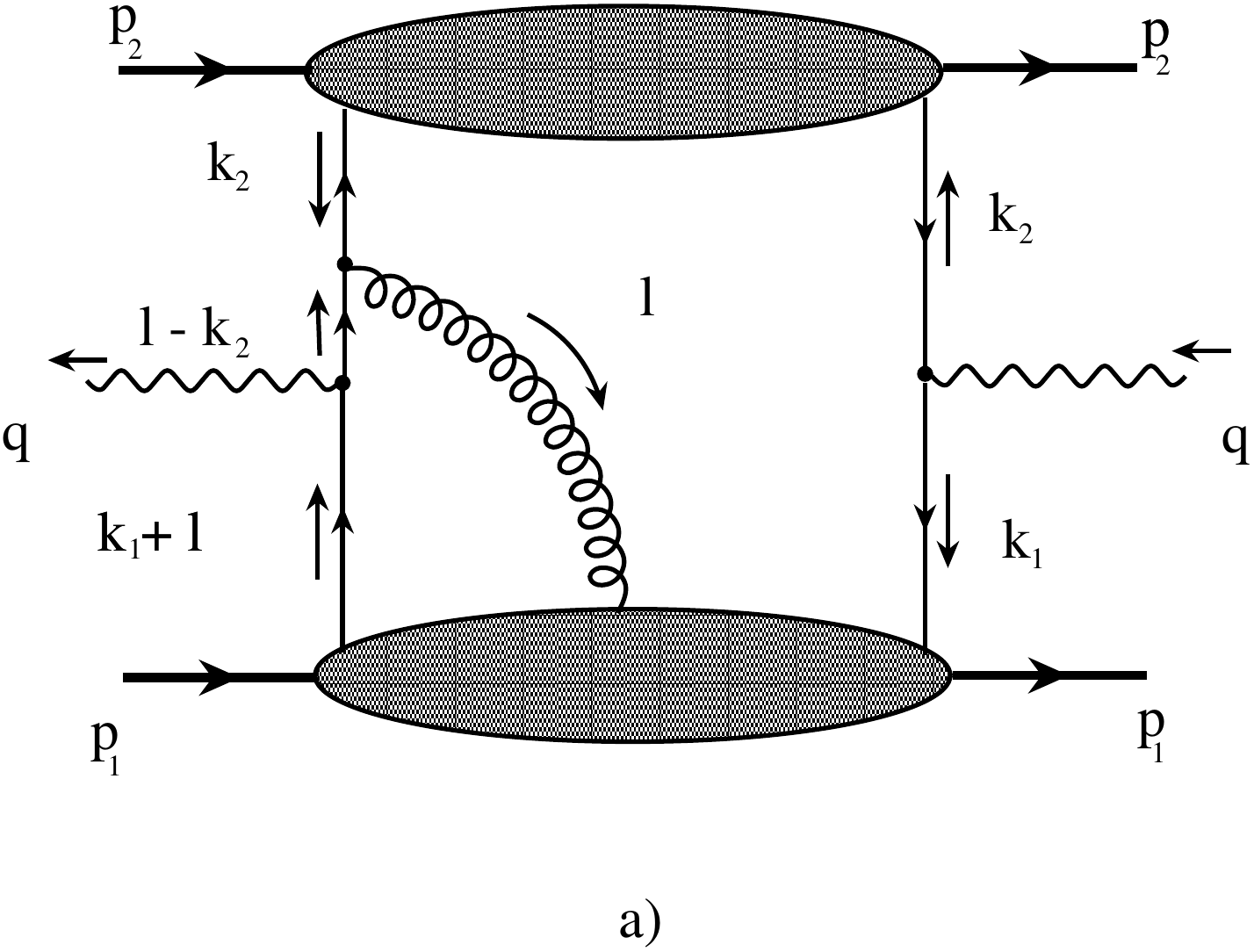}
\hspace{1.cm}\includegraphics[width=0.3\textwidth]{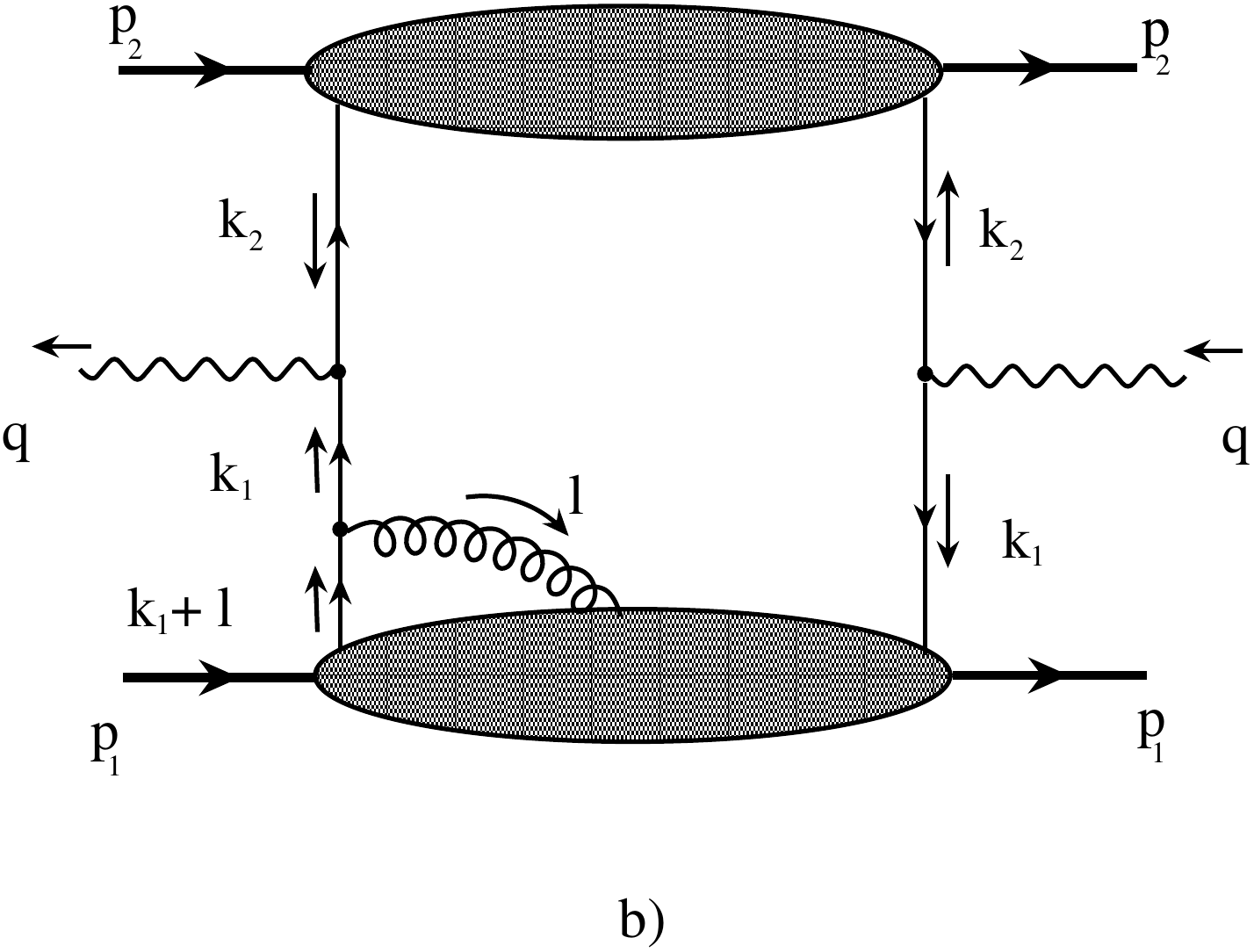}}
%\vspace{-2.5cm}
\caption{The Feynman diagrams which contribute to the polarized Drell-Yan hadron tensor.}
\label{Fig-DY}
\end{figure}

For pedagogical reasons, we remind briefly our findings for the Drell-Yan (DY) process where
one of hadrons possesses
the transverse polarization, see \cite{AT-10} for all details.
As usual, the Drell-Yan process with the transversely polarized nucleon is defined as
\begin{eqnarray}
\label{DY}
N^{(\uparrow\downarrow)}(p_1) + N(p_2) \to \gamma^*(q) + X(P_X)\to
\ell(l_1) + \bar\ell(l_2) + X(P_X)\,,
\end{eqnarray}
where the virtual photon producing the lepton pair ($l_1+l_2=q$) has a large mass squared
($q^2=Q^2$)
while the transverse momenta are small and integrated out.
The dominant light-cone directions for the DY process (Fig.~\ref{Fig-DY})
are defined as
$p_1\approx n^*\,Q/(x_B \sqrt{2})$ and $p_2\approx n \,Q/(y_B \sqrt{2})$  with
the dimensionless light-cone vectors
$n^*_{\mu}=(n^{*\,+},\, 0^-,\, {\bf 0}_\perp)$ and $ n_{\mu}=(0^{+},\,n^{-} ,\, {\bf 0}_\perp)$.

Since we deal with a large $Q^2$ in the process under consideration, it is possible to apply the
factorization theorem to get the corresponding hadron tensor factorized in the form of convolution:
\begin{eqnarray}
\label{Fac-DY}
\text{Hadron tensor} = \{\text{Hard part (pQCD)}\} \otimes
\{\text{Soft part (npQCD)} \}\,.
\end{eqnarray}
Usually both the hard and soft parts in Eqn. (\ref{Fac-DY}) are independent of each other,
UV- and IR-renormalizable. Moreover, various parton distributions which parametrize the soft part
have to manifest the universality property.

Based on DY-process, it is convenient to study the role of twist $3$ by exploring
of different kinds of asymmetries, for instance,
the left-right asymmetry. This left-right asymmetry means the transverse momenta
of the leptons are correlated with the direction
$\textbf{S}\times \textbf{e}_z$ where $S_\mu$ implies the
transverse polarization vector of the nucleon and $\textbf{e}_z$ is a beam direction \cite{Barone}.
Generally speaking, any single spin asymmetries (SSAs) can be presented in the following symbolical form
(at this moment, the exact expression for SSA is irrelevant):
\begin{eqnarray}
{\cal A} \sim d\sigma^{(\uparrow)} - d\sigma^{(\downarrow)}
\sim {\cal L}_{\mu\nu}\, H_{\mu\nu}\, ,
\end{eqnarray}
where ${\cal L}_{\mu\nu}$ is an unpolarized leptonic tensor and, consequently, has only the real part;
$H_{\mu\nu}$ stands for the hadronic tensor which is also real.
Since one of hadrons is transversely polarized,
the corresponding matrix element which forms the soft part of hadron tensor reads \cite{AT-10}
\begin{eqnarray}
\label{parVecDY}
\langle p_1, S^T | \bar\psi(\lambda_1 \tilde n)\, \gamma^+ \,
g A^{\alpha}_T(\lambda_2\tilde n) \,\psi(0)
|S^T, p_1 \rangle = i\varepsilon^{\alpha + S^T -} (p_1p_2)
\int dx_1 dx_2 \, e^{i x_1\lambda_1+ i(x_2 - x_1)\lambda_2}
\, B^V(x_1,x_2)\, ,
\end{eqnarray}
where the light-cone vector $\tilde n$ is a dimensionful analog of vector $n$.
Therefore, in order to provide for the condition, $H_{\mu\nu}\in\Re\text{e}$,
the complex $i$ in the {\it r.h.s.} of (\ref{parVecDY})
has to be compensated either (a) by the complexness of the hard part (this is the standard contribution):
\begin{eqnarray}
\label{HadTen-a}
H^{(a)}_{\mu\nu} \sim i\,\Im\text{m}\,[ \text{Hard}] \otimes
\biggl\{ \langle p_1,S_T|{\cal O}(\bar\psi,\psi,A) |S_T,p_1 \rangle
\stackrel{{\cal F}}{\sim} \, i \varepsilon^{\alpha + S_T -} B^V
\biggr\}  \, ,
\end{eqnarray}
or (b) by the complexness of the soft part
\begin{eqnarray}
\label{HadTen-b}
H^{(b)}_{\mu\nu} \sim  \text{Hard} \otimes
\biggl\{  \langle p_1,S_T|{\cal O}(\bar\psi,\psi,A) |S_T,p_1 \rangle
\stackrel{{\cal F}}{\sim} \, i \varepsilon^{\alpha + S_T -}
\,i\,\Im\text{m}\,[ B^V ]
\biggr\}  \, .
\end{eqnarray}
In Eqns. (\ref{HadTen-a}) and (\ref{HadTen-b}), $\stackrel{{\cal F}}{\sim}$ and ${\cal O}(\bar\psi,\psi,A)$
are the shorthands for the Fourier transformation and the corresponding quark-gluon operator, respectively
(Eqn. (\ref{parVecDY})). In general, the hard parts in Eqns.
(\ref{HadTen-a}) and (\ref{HadTen-b}) differ from each other. For instance, the hard part of the diagram
presented in Fig.~\ref{Fig-DY}(a) contains the quark propagator in contrast to the hard part of
the diagram in Fig.~\ref{Fig-DY}(b)
\footnote{The corresponding $\delta$-functions appeared in the hadron tensor
and expressed the momentum conservation low should be also referred to the hard parts.
This statement was argued in \cite{An} in the context of the so-called factorization links.}.

However, in the previous studies (for example, \cite{Boer:2003cm, BQ, Teryaev, Boer}),
$B^V(x_1,x_2)$-function has been assumed to be a purely real function:
\begin{eqnarray}
\label{g-pole-B}
B^V(x_1,x_2)= \frac{{\cal P}}{x_1-x_2} T(x_1,x_2)\,
\end{eqnarray}
with the function $T(x_1,x_2)\in\Re\text{e}$ which
parametrizes the corresponding projection of $\langle \bar\psi\, G_{\alpha\beta}\,\psi \rangle$.
Therefore, the scenario (b) (Eqn. (\ref{HadTen-b})) will never be realized if the $B^V(x_1,x_2)$-function
has the form as in Eqn. (\ref{g-pole-B}).
As a result, the QED gauge invariance of the DY hadron tensor is in question.
Indeed, having analyzed the hard subprocess in the context of the QED gauge invariance,
we can conclude that only the sum of two diagrams in Fig.~\ref{Fig-DY} (a) and (b)
ensures the QED gauge invariance of the hadron tensor. As shown in \cite{AT-10},
the contribution of $H^{(a)}_{\mu\nu}$ is associated with the Feynman diagram in
Fig.\ref{Fig-DY}(a) while the contribution of $H^{(b)}_{\mu\nu}$ is
generated by the diagram presented in Fig.~\ref{Fig-DY}(b).
%%%%%%%%%%%%%%%%%%%%%%%%%%%%% FIGURE %%%%%%%%%%%%%%%%%%%%%%%%%%%%%%%%
%%%%%%%%%%%%%%%%%%%%%%%%%%%%%%%%%%%%%%%%%%%%%%%%%%%%%%%%%%%%%%%%%%%%%%%

Thus, we may infer that the real function $B^V(x_1,x_2)$
in the form as in Eqn. (\ref{g-pole-B})
leads to the {\it violation} of the QED gauge invariance of the hadron tensor.

To solve this discrepancy, it is instructive to remind the reason which leads to the
representation in Eqn. (\ref{g-pole-B}).
%In this work we will work within the axial-type gauge.
The conclusion that $B^V$ is a real function has come from
the ambiguity in the solutions of the differential equation (provided that $A^+=0$):
\begin{eqnarray}
\label{DiffEqnsG}
\partial^+\, A^\alpha_T=G^{+\,\alpha}_T\, .
\end{eqnarray}
Indeed, the formal solutions of Eqn. (\ref{DiffEqnsG}) have the following forms:
\begin{eqnarray}
\label{Ag1}
A^\mu(z)&=&
\int\limits_{-\infty}^{\infty} d\omega^-
\theta(z^- - \omega^-) G^{+\mu} (\omega^-)
+ A^\mu(-\infty) \,
\\
\label{Ag2}
&=&
- \int\limits_{-\infty}^{\infty} d\omega^-
\theta(\omega^- - z^- ) G^{+\mu} (\omega^-)
+ A^\mu(\infty) \, .
\end{eqnarray}
From the first glance, Eqns. (\ref{Ag1}) and (\ref{Ag2}) seem to be equivalent each other.
However, as we will see below, this is not true.

Now, if we insert Eqns. (\ref{Ag1}) and (\ref{Ag2}) into Eqn. (\ref{parVecDY}) and use the following parametrization
\begin{eqnarray}
\label{parT}
\langle p_1, S^T | \bar\psi(\lambda_1 \tilde n)\, \gamma^+ \,
\tilde n_\nu G^{\nu\alpha}_T(\lambda_2\tilde n) \,\psi(0)
|S^T, p_1 \rangle=
\varepsilon^{\alpha + S^T -}\,(p_1p_2)\,
\int dx_1 dx_2 \, e^{i x_1\lambda_1+ i(x_2 - x_1)\lambda_2} \, T(x_1,x_2)\, ,
\end{eqnarray}
we get the following representations:
\begin{eqnarray}
\label{Phi1}
B^V(x_1,x_2)&=&\delta(x_1-x_2)B^V_{A(-\infty)}(x_1)+
\frac{T(x_1,x_2)}{x_1-x_2 + i\epsilon}\,,
\\
\label{Phi2}
&=&\delta(x_1-x_2)B^V_{A(+\infty)}(x_1)+
\frac{T(x_1,x_2)}{x_1-x_2 - i\epsilon}\,,
\end{eqnarray}
respectively.
In Eqns. (\ref{Phi1}) and (\ref{Phi2}), the corresponding prescriptions $\pm i\epsilon$ arise from the
integral representation for the theta-function:
\begin{eqnarray}
\label{theta-func}
\theta(\pm x) =\frac{\pm i}{2\pi} \int\limits_{-\infty}^{+\infty} dk\, \frac{e^{-ikx}}{k\pm i\epsilon}\,.
\end{eqnarray}

Further, if we suppose that the representations (\ref{Phi1}) and (\ref{Phi2}) are equivalent to each other,
we can calculate the plus and minus combinations of (\ref{Phi1}) and (\ref{Phi2}) resulting in
\begin{eqnarray}
B^V(x_1,x_2)&=&\frac{1}{2}B^V(x_1,x_2) + \frac{1}{2}B^V(x_1,x_2)
\\
&=&
 \frac{1}{2}\delta(x_1-x_2) \Big\{
B^V_{A(-\infty)}(x_1)+B^V_{A(+\infty)}(x_1)
\Big\} +
\frac{{\cal P}}{x_1-x_2}T(x_1,x_2)\,
\nonumber
\end{eqnarray}
and
\begin{eqnarray}
0&=&B^V(x_1,x_2)-B^V(x_1,x_2)
\\
&=&
\delta(x_1-x_2) \Big\{
B^V_{A(+\infty)}(x_1)-B^V_{A(-\infty)}(x_1)
\Big\} -
2i\,\pi\, \delta(x_1-x_2)T(x_1,x_2)\,.
\nonumber
\end{eqnarray}
The ambiguity in the solutions of (\ref{DiffEqnsG}) ((\ref{Phi1}) and (\ref{Phi2}))
ultimately gives us the standard representation (\ref{g-pole-B}) with the real function $B^V$
provided the asymmetric boundary condition for gluons is given by
$B^V_{A(\infty)}(x) = - B^V_{A(-\infty)}(x)$.

In fact, the representations (\ref{Phi1}) and (\ref{Phi2}) are {\it not} equivalent ones \cite{AT-10}.
To see that it is necessary to remember that all axial-type gauges are the particular cases of the most general
contour gauge (see Appendix \ref{ContourGauge:App:A} and the text below for details).
Using the contour gauge conception, one can easily check that
the representation (\ref{Phi1}) belongs to the gauge $[x,\,-\infty]=1$, while
the representation (\ref{Phi2}) belongs to the gauge $[+\infty,\, x]=1$. Therefore, one has {\it no any
reason} to believe that (\ref{Phi1}) and (\ref{Phi2}) are equivalent ones (for details, see (\ref{B-plus}), (\ref{B-minus})),{\it i.e.}
\begin{eqnarray}
\Big\{ \text{Eq.(\ref{Phi1})}\Longrightarrow B^V_{+}(x_1,x_2)\Big\} \quad \not= \quad
\Big\{ B^V_{-}(x_1,x_2) \Longleftarrow \text{Eq.(\ref{Phi2})}\Big\}\,.
 \end{eqnarray}
Roughly speaking, it resembles the trivial situation where two different vectors have the same projection on the
certain direction. In this context, the well-known axil gauge $A^+=0$ can be treated as some type of ``projection" which
corresponds to two different ``vectors" represented by two different contour gauges (see Appendix \ref{ContourGauge:App:A}).

We now consider the QED gauge invariant hadron tensor for the DY-process with the transverse
polarization. In order to understand which contour gauge we need to deal with, before imposing the condition $A^+=0$,
it is necessary to take into account the contributions of
$\langle p_1,S^T| \bar\psi\, \gamma^+\,  A^+\,\psi |S^T,p_1\rangle$
in the standard hadron tensor (see Fig.~\ref{Fig-DY}(a)).
Based on the analysis of the $\gamma$-structure of this diagram as shown in \cite{AT-10},
the Feynman causal prescription in the quark propagator (see, eqns. (2), (6) and (8) of \cite{AT-10})
uniquely leads to the Wilson line in the quark-gluon correlator:
\begin{eqnarray}
\label{Pexp-1}
[-\infty^-,\, 0^-] = Pexp\Big\{ - i g \int\limits_{-\infty}^{0}
 dz^- \, A^+(0,z^-,\vec{{\bf 0}}_T) \Big\}\,.
\end{eqnarray}
Eqn. (\ref{Pexp-1}) suggests that we have to use the contour gauge $[-\infty^-,\, 0^-]=1$. Therefore, the contour gauge
defined by $[-\infty^-,\, 0^-]=1$ destroys the ambiguity, and
the function $B^V(x_1,x_2)$ has to be described by the following representation (see, (\ref{B-plus}) and \cite{AT-10} for details)
\begin{eqnarray}
\label{B-plus0}
&&B^V_{+}(x_1,x_2)=
\frac{T(x_1,x_2)}{x_1-x_2 + i\epsilon}\,.
\end{eqnarray}
As a result,
the diagram represented in Fig.~\ref{Fig-DY}(b) (or the hadron tensor (\ref{HadTen-b}))
does contribute to the hadron tensor, and this together with the first diagram represented in Fig.\ref{Fig-DY}(a) form
the gauge invariant (GI) hadron tensor (see, Eqn. (34) of \cite{AT-10}):
\begin{eqnarray}
\label{HadTen-GI}
\overline{\cal W}_{\text{GI}}^{\mu\nu}=
\overline{\cal W}_{(1)}^{\mu\nu} + \overline{\cal W}_{(2)}^{\mu\nu} =
- \frac{2}{q^2}\,\varepsilon^{\nu S^T p_1 p_2} \, \big(x_B \, p_{1}^{\mu}- y_B\, p_{2}^{\mu}\big)
\, \bar q(y_B)\, T(x_B,x_B) \, .
\end{eqnarray}
From Fig.~\ref{Fig-DY}, we can also realize that the representation of $B^V(x_1,x_2)$ with the complex prescription
$+i\,\epsilon$ in the gluonic pole, see (\ref{B-plus}), corresponds to
the initial state interaction (ISI) with respect to the hard subprocess.
We want to stress that in the case of DY-process we deal with the initial state interaction only
as opposed to the case of the direct photon production which is studied below.

For the DY-process, the initial state interaction generates $-2\ell^+ k_2^- + i\epsilon$
(see, the diagram in Fig.~\ref{Fig-DY}(a))
in the quark propagator which, in turn, leads to (i)
the contour gauge $[-\infty^-,\, 0^-]=1$ and, then, to (ii)
the function $B^V_+$ with the certain complex prescription (\ref{B-plus}). The latter ensures
the QED gauge invariance for the hadron tensor. Schematically, the mentioned logical chain
can be presented as
\begin{eqnarray}
\label{DY-illust}
\text{\bf ISI}\Rightarrow \frac{1}{-\ell^+ + i\epsilon} \Rightarrow \text{gauge}\,\,\,[z^-,\, -\infty^-]=1
\Rightarrow \frac{T(x_1,x_2)}{x_1-x_2 +i\epsilon} \Rightarrow\text{\bf GI}\,.
\end{eqnarray}
We can see that the prescription in the quark propagator of the hard part
gives information on the contour gauge for gluons from the soft part. In other words
 the hard and soft parts are not fully independent of each other.
Despite of this, the DY hadron tensor has formally been factorized with the mathematical convolution,
and the parton distributions, such as the twist-3 function $B^V(x_1,x_2)$, still satisfy the universality condition.
In contrast to the DY-process, as we will see in the next section, the direct photon production tensor
is built with the functions $B^V(x_1,x_2)$ that will not manifest the universality.

We will refer to {\it a soft breaking} of factorization
when the factorization procedure results in the
mathematical convolution between the finite hard and soft parts, but
there is no universality for the soft functions or the hard and soft parts are not
totally independent.

To conclude this section, let us note that the situation with
the gauge invariance for the DY-process (\ref{DY-illust}) is very similar
to what has been discussed for the vector meson electroproduction in \cite{Braun}
where the final state interaction also predetermined the correct prescription for the spurious gluon pole for the validity
of QCD factorization.

\section{Hadron tensor of the direct photon production I:
kinematics and gauge invariance}

\subsection{Kinematics}

In this section, we study the two hadron collisions,
where one of the hadrons possesses the transverse polarization,
which produce the direct photon in the final state in:
\begin{eqnarray}
\label{process}
N^{(\uparrow\downarrow)}(p_1) + N(p_2) \to \gamma(q) + q(k) + X(P_X)\,.
\end{eqnarray}
The gluonic poles are being manifested in this process in the similar way as for the Drell-Yan process \cite{Teryaev}.
We perform our calculations within a {\it collinear} factorization, and
it is convenient (see, e.g., \cite{An})
to fix the dominant light-cone directions as
\begin{eqnarray}
\label{kin-DY}
&&p_1 = \sqrt{\frac{S}{2}}\, n^*\, ,
\quad p_2 = \sqrt{\frac{S}{2}}\, n\,,\quad \text{with}
\nonumber\\
&&
n^*_{\mu}=(1/\sqrt{2},\,{\bf 0}_T,\,1/\sqrt{2}), \quad n_{\mu}=(1/\sqrt{2},\,{\bf 0}_T,\,-1/\sqrt{2})\, .
\end{eqnarray}
The hadron momenta $p_1$ and $p_2$ have the plus and minus dominant light-cone
components, respectively. Accordingly, the quark and gluon momenta $k_1$ and $\ell$ lie
along the plus dominant direction while the gluon momentum $k_2$ -- along the minus direction.
The final on-shell photon and quark(anti-quark) momenta can be presented as
\begin{eqnarray}
\label{Photon-Quark}
&&q = y_B\, \sqrt{\frac{S}{2}}\, n - \frac{q_\perp^2}{y_B \sqrt{2 S}}\, n^* + q_\perp\,,
\nonumber\\
&&k = x_B\, \sqrt{\frac{S}{2}}\, n^* - \frac{k_\perp^2}{x_B \sqrt{2 S}}\, n + k_\perp\,.
\end{eqnarray}
The Mandelstam variables for the process and subprocess are defined as
\begin{eqnarray}
\label{MandVar}
&&S=(p_1+p_2)^2, \quad T=(p_1-q)^2,\quad U=(q-p_2)^2,
\nonumber\\
&& \hat s =(x_1p_1 + yp_2)^2 = x_1 y S, \quad \hat t=(x_1p_1 -q)^2=x_1 T, \quad
\hat u =(q-yp_2)^2 = yU.
\end{eqnarray}

\begin{figure}[t]
\centerline{
\includegraphics[width=0.3\textwidth]{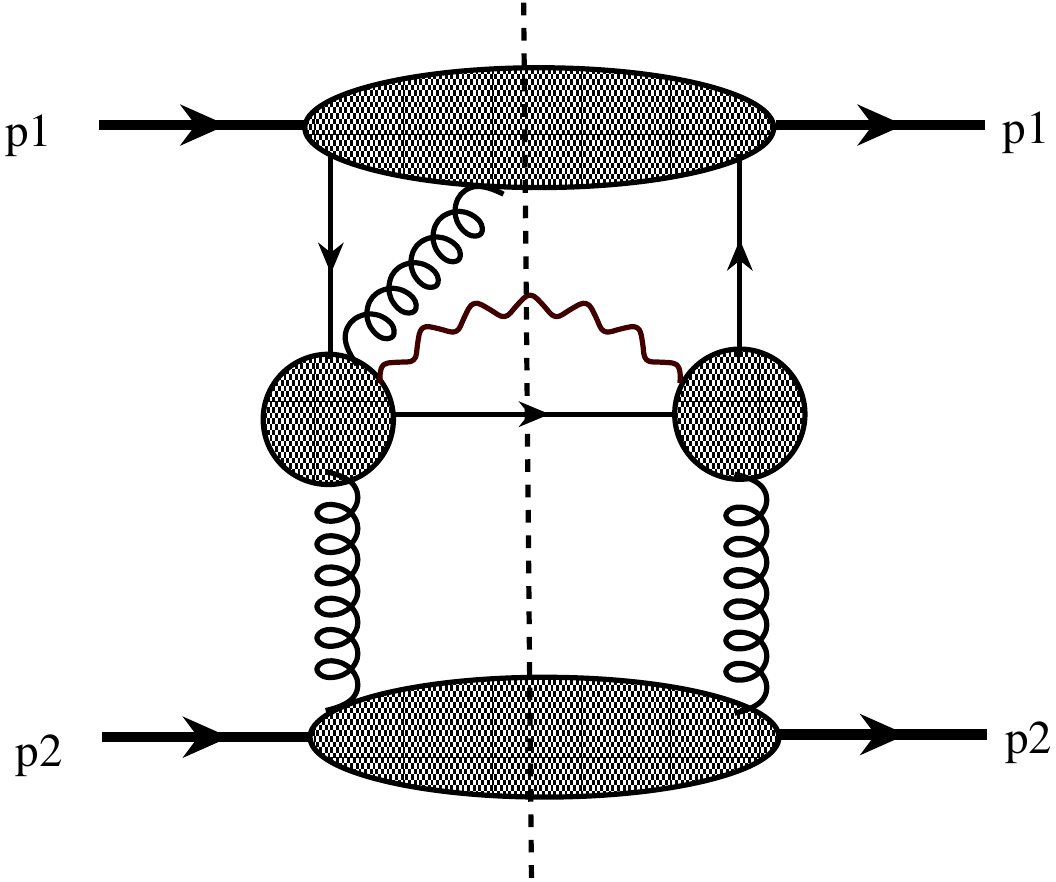}}
\caption{The Feynman diagram describing the hadron tensor of the
direct photon production.}
\label{Fig-DirPhot}
\end{figure}

The amplitude of process (\ref{process}) involves the contributions from
\begin{itemize}
\item the leading (LO) diagrams: two diagrams with a radiation of the photon before (${\cal A}^{\text{LO}}_1$)
and after (${\cal A}^{\text{LO}}_2$)
the quark-gluon vertex with the gluon going to the lower blob, see the right side of Fig.~\ref{Fig-DirPhot};
\item the next-to-leading order (NLO) diagrams: eight diagrams constructed from
the LO diagrams by insertion of all possible radiations
of the additional gluon which together with the quark goes to the upper blob,
see the left side of Fig.~\ref{Fig-DirPhot}.
\end{itemize}
We denote the sum of diagrams as
\begin{eqnarray}
\label{amp}
{\cal A}^{\text{LO}}_1 + {\cal A}^{\text{LO}}_2 + \sum\limits_{i=1}^{8}{\cal B}^{\text{NLO}}_i\,
\end{eqnarray}
that generates the hadron tensor related to the corresponding asymmetry:
\begin{eqnarray}
\label{SSA}
d\sigma^\uparrow - d\sigma^\downarrow \sim {\cal W}=
\sum\limits_{i=1}^{2}\sum\limits_{j=1}^{8}
{\cal A}^{\text{LO}}_i \ast {\cal B}^{\text{NLO}}_j\,.
\end{eqnarray}
In this work, we mainly dwell on the discussion of the hadron tensor rather than the asymmetry
itself. Diagrammatically (Fig.~\ref{Fig-DirPhot}), the hadron tensor can be presented
in the form of an interference between the LO and NLO diagrams:
${\cal A}^{\text{LO}}_i \ast {\cal B}^{\text{NLO}}_j$.
In Fig.~\ref{Fig-DirPhot}, the upper blob determines the matrix element of the twist-3 quark-gluon operator
while the lower blob -- the matrix element of the twist-2 gluon operator related to the
unpolarized gluon distribution.
%%%%%%%%%%%%%%%%%%%%%%%%%%%%% FIGURE %%%%%%%%%%%%%%%%%%%%%%%%%%%%%%%%
%%%%%%%%%%%%%%%%%%%%%%%%%%%%%%%%%%%%%%%%%%%%%%%%%%%%%%%%%%%%%%%%%%%%%%%

\subsection{Factorization procedure}

Since the collinear factorization is our main tool, let us outline the main stages of factorization.
The factorization procedure contains the following steps:
\begin{itemize}
\item the decomposition of loop integration momenta around the corresponding dominant direction:
$k_i = x_i p + (k_i\cdot p)n + k_T$
within the certain light cone basis formed by the vectors $p$ and $n$ (in our case, $n^*$ and $n$);
\item the replacement:
$d^4 k_i \Longrightarrow d^4 k_i \,dx_i \delta(x_i-k_i\cdot n)$
that introduces the fractions with the appropriated spectral properties;
\item the decomposition of the corresponding propagator products around the dominant direction:
\begin{eqnarray}
H(k) = H(xp) + \frac{\partial H(k)}{\partial k_\rho} \biggl|_{k=xp}\biggr. \, k^T_\rho + \ldots \,;
\nonumber
\end{eqnarray}
\item the use of the collinear Ward identity, if it requests by the needed approximation:
\begin{eqnarray}
\frac{\partial H(k)}{\partial k_\rho} = H_{\rho}(k,\,k) \, ;
\nonumber
\end{eqnarray}
\item performing of the Fierz decomposition for $\psi_\alpha (z) \, \bar\psi_\beta(0)$ in
the corresponding space up to the needed projections.
\end{itemize}
Notice that, for our purposes, it is enough to be limited by the first order of decomposition in the third item.
As a result of this procedure, we should reach the factorized form for the considered subject, see (\ref{Fac-DY}).

\subsection{Hadron tensor: QED gauge invariance}

At the first item, we want to discuss the QED gauge invariance of the hadron tensor.
To check the QED gauge invariance, it is sufficient to consider
the typical Feynman diagrams H1, H3 and H5 represented in Fig. \ref{Fig-All}.
For the definiteness, we pay our attention on the anti-quark contribution.
All our results can trivially be extended to the quark contribution as well.

Before factorization,
the H1-diagram in Fig.\ref{Fig-All} leads to the following expression:
\begin{eqnarray}
\label{HT-dia-1}
&&\overline{{\cal W}}(\text{diag.H1})=\int (d^4k_1)\,(d^4k_2) \, \delta^{(4)}(k_1+k_2-k-q) \,
\Phi_g^{\alpha\beta}(k_2)\, \text{C}_2 \times
\nonumber\\
&&\bar v(k)\hat\varepsilon S(k+q)
\gamma^\alpha_\perp \gamma^- \gamma^\beta_\perp
\int (d^4\ell)\, S(\ell+k+q)\gamma^\rho_\perp S(k+q) \hat q v(k)\,
\bar\Phi_\perp^{[\gamma^+],\rho}(k_1,\ell)\,,
\end{eqnarray}
where $\text{C}_2$ implies the corresponding colour factor.
Here and in what follows the coupling constants are not shown explicitly.
In Eqn. (\ref{HT-dia-1}) the gluon (unpolarized) twist-2 parameterizing function is defined as
\begin{eqnarray}
\label{g-func}
\Phi_g^{\alpha\beta}(k_2)= \int\hspace{-0.5cm}\sum\limits_X\,\int (d^4\eta_2)\, e^{-ik_2\eta_2}\,
\langle p_2| A^\alpha(0)|P_X\rangle \, \langle P_X | A^\beta(\eta_2) | p_2\rangle =
g^{\alpha\beta}_\perp\, {\cal F}^g(k_2)
\end{eqnarray}
while the quark-gluon twist-3 parameterizing function is given by
\begin{eqnarray}
\label{q-func}
\bar\Phi_\perp^{[\gamma^+],\rho}(k_1,\ell)= \int\hspace{-0.5cm}\sum\limits_X\,\int (d^4\eta_1)\,(d^4z)\,
 e^{-ik_1\eta_1-i\ell z}\,\rm{tr}_{\it D}\,
\langle p_1, S^T| \psi(\eta_1) |P_X\rangle \, \langle P_X | \bar\psi(0)\gamma^+ A^\rho_\perp(z) | S^T, p_1\rangle\,.
\end{eqnarray}
We now carry out the standard factorization procedure and, after some algebra, obtain the following expression:
\begin{eqnarray}
\label{HTFac-dia-1}
&&\overline{{\cal W}}(\text{diag.H1})= -2 \,\int dx_1\, dy \, \delta^{(4)}(x_1 p_1+y p_2-k-q) \,
{\cal F}^g(y)\, \text{C}_2 \times
\nonumber\\
&&\bar v(k)\hat\varepsilon \, \frac{\gamma^+}{2x_1 p_1 + i\epsilon}
 \gamma^- \,
\int dx_2\,  \frac{\gamma^+}{2x_2 p_1 + i\epsilon} \,
\gamma^\rho_\perp \frac{\gamma^-}{2y p_2 + i\epsilon} \hat q v(k)\,
\bar\Phi_\perp^{[\gamma^+],\rho}(x_1,x_2)\,,
\end{eqnarray}
where
\begin{eqnarray}
\label{q-func-fac}
\bar\Phi_\perp^{[\gamma^+],\rho}(x_1,x_2)&=& \int\hspace{-0.5cm}\sum\limits_X\, \int (d\lambda_1)\,(d\lambda_2)\,
 e^{-ix_1\lambda_1-i(x_2-x_1)\lambda_2}\,
\rm{tr}_{\it D}\,
\langle p_1, S^T| \psi(\lambda_1 n) |P_X\rangle \, \langle P_X | \bar\psi(0)\gamma^+ A^\rho_\perp(\lambda_2 n) | S^T, p_1\rangle\,
\nonumber\\
&=&\varepsilon^{\rho + S^T -} (p_1p_2) \, B^V(x_1,x_2).
\end{eqnarray}

In the diagrams  H1, H3 and H5, we deal with the final state interaction (FSI)
with respect to the hard part. Therefore,
the $B^V$-function has the representation as in (\ref{B-minus}) (cf. \cite{AT-10}), {\it i.e.}
\begin{eqnarray}
\label{B-fsi}
B^V_{-}(x_1,x_2)\equiv B^V_{\text{FSI}}(x_1,x_2)=\frac{T(x_1,x_2)}{x_1 - x_2 - i\epsilon}\, .
\end{eqnarray}
In the similar way as in the preceding section,
we have to restore the path in the Wilson line with $A^+$-fields in the quark-gluon correlators
which appear in the diagrams H1, H3 and H5 in Fig.\ref{Fig-All}.
After straight-forward calculations, we derive the Wilson line in the form
$[+\infty^-,\, z^-]$ that suggests us to use the representation (\ref{B-minus}) for our $B^V$-function.

The contribution of H5-diagram in Fig.\ref{Fig-All} is equal to zero due to
the fact that the photon momentum has the dominant minus light-cone component and, therefore,
the $\gamma$-structure gives $(\gamma^-)^2=0$.

We now calculate the hadron tensor term associated with the H3-diagram in Fig.\ref{Fig-All},
it reads
\begin{eqnarray}
\label{HTFac-dia-3}
&&\overline{{\cal W}}(\text{diag.H3})= -2 \,\int dx_1\, dy \, \delta^{(4)}(x_1 p_1+y p_2-k-q) \,
{\cal F}^g(y)\, \text{C}_2 \times
\nonumber\\
&&\bar v(k)\hat\varepsilon \, \frac{\gamma^+}{2x_1 p_1 + i\epsilon}
 \gamma^-  \hat q \,
\int dx_2\,  \frac{\gamma^-}{- 2y p_2 + i\epsilon}
 \frac{\gamma^+}{2x_2 p_1 + i\epsilon} \gamma^\rho_\perp v(k)\,
\bar\Phi_\perp^{[\gamma^+],\rho}(x_1,x_2)\,,
\end{eqnarray}
where the $B^V$-function is also given by the representation (\ref{B-fsi}) or (\ref{B-minus}).

After some $\gamma$-algebra, we can check that the contribution of (\ref{HTFac-dia-3}) is equal to
the contribution of (\ref{HTFac-dia-1}) but with an opposite sign. Therefore, the sum of all
contribution gives us zero (we remind that while the second diagram contribution is equal to zero itself)
{\it i.e.}
\begin{eqnarray}
\label{QED-GI}
\overline{{\cal W}}(\text{diag.H1})+\overline{{\cal W}}(\text{diag.H3})+\overline{{\cal W}}(\text{diag.H5})=0\,.
\end{eqnarray}
In fact, the identity (\ref{QED-GI}) reflects the QED gauge invariance for the hadron tensor.
We emphasize that the QED gauge invariance takes place owing to
the same complex prescriptions in the definitions of $B^V$-function (Eqn. \ref{B-fsi}).
In turn, the same complex prescriptions emanate from
the final state interaction presented in H1, H3 and H5 of Fig.\ref{Fig-All}.
Finally, using the similar logical chain as for the DY-process, we can write
that
%for the QED gauge invariance of the hadron tensor for the direct photon production we have
\begin{eqnarray}
\label{DPP-illust}
\text{\bf FSI}\Rightarrow \frac{1}{\ell^+ + i\epsilon} \Rightarrow \text{gauge}\,\,\,[+\infty^-,\, z^- ]=1
\Rightarrow \frac{T(x_1,x_2)}{x_1-x_2 - i\epsilon} \Rightarrow\text{\bf QED GI}\,.
\end{eqnarray}

We would like to emphasize that the concrete sign of the gluonic pole prescription is not so crucial
 for the QED gauge invariance
 because here we deal with only one type of interaction which is
the final state interaction (this distinguishes the case of QCD gauge invariance
which is considered below). It is more important to have the same
prescriptions in all gluonic poles. In other words, from the point of view  of contour gauge,
we may use the same wrong $+i\epsilon$ prescription in
the representation of $B^V$-function for diagrams H1, H3 and H5 depicted in Fig.\ref{Fig-All}.
But this wrong prescription still leads to the QED gauge invariance.
However, after calculation of the
imaginary parts for the corresponding asymmetry, the wrong prescription definitely plays a negative role.

\subsection{Hadron tensor: QCD gauge invariance}

We are now in a position to dwell on the QCD gauge invariance of the hadron tensor
for the direct photon production. To check this invariance,
we have to consider four typical diagrams H1, H5, D1 and H9, depicted in Fig.\ref{Fig-All}, which come
from the corresponding $\xi$-process (see, \cite{BogoShir}). Notice that the gluon
enters in the quark-gluon correlator as an internal field. For the QCD gauge invariance, we have to assume
that all charged particles are on-shell, {\it i.e} we deal with the physical gluons only.
The substantial differences between this case and a pure perturbative Compton scattering case are discussed in Appendix B.

To write down the Ward identity, we need to replace the gluon transverse polarization $\epsilon^T_\alpha$
by the gluon longitudinal momentum $\ell^L_\alpha$ in the quark-gluon correlator:
\begin{eqnarray}
\label{qg-col-Rep}
&&\bar\Phi_\perp^{[\gamma^+],\rho}(k_1,\ell)=
- \int (d^4\eta_1)\,
 e^{-ik_1\eta_1}\, \epsilon_T^\rho\,
\langle p_1, S^T| \bar\psi(0)\gamma^+ \psi(\eta_1)\, a^+(\ell)| S^T, p_1\rangle
\stackrel{\epsilon^T\rightarrow \ell^L}{\Longrightarrow}
\nonumber\\
&&- \int (d^4\eta_1)\,
 e^{-ik_1\eta_1}\, \ell_L^\rho\,
\langle p_1, S^T| \bar\psi(0)\gamma^+ \psi(\eta_1)\, a^+(\ell)| S^T, p_1\rangle\,,
\end{eqnarray}
where $a^+(\ell)$ stands for the gluon creation operator. The summation over the intermediate states
is not shown explicitly. Notice that the parametrization of this correlator
through $B^V$-function stays unchanged.

Consider now the contribution of the H1-diagram in Fig.\ref{Fig-All} to the hadron tensor.
Before going further, it is instructive to begin with
the gluon loop integration corresponding to the mentioned diagram. We have
\begin{eqnarray}
\label{gluon-loop-1}
\int (d^4\ell) S(\ell+k+q) \hat\ell_L \langle ... a^+(\ell) ...\rangle\,,
\end{eqnarray}
where we do not explicitly  write the operators which are irrelevant at the moment (cf. (\ref{qg-col-Rep})).
After factorization, we obtain
\begin{eqnarray}
\label{gluon-loop-2}
&&\int dx_2 \int (d^4\ell) \delta(x_2-x_1-\ell n) S(\ell+k+q) \hat\ell_L \langle ... a^+(\ell) ...\rangle=
\nonumber\\
&&\int dx_2  S(x_2 p_1+yp_2) \, (x_2-x_1)\hat p_1 \,
\int (d^4\ell) \delta(x_2-x_1-\ell n) \langle ... a^+(\ell) ...\rangle
\,,
\end{eqnarray}
where we decompose the hard part around the dominant direction  and
put $\ell_L=(x_2-x_1)p_1$ which is actually dictated by the $\gamma$-structure
and the momentum conservation. Using all these, we get the following expression:
\begin{eqnarray}
\label{QCD-WI-1}
&&\overline{{\cal W}}(\text{diag.H1})=
-2 \,\int dx_1\, dy \, \delta^{(4)}(x_1 p_1+y p_2-k-q) \,
{\cal F}^g(y)\, \text{C}_2 \times
\nonumber\\
&&\bar v(k)\hat\varepsilon \, \frac{\gamma^+}{2x_1 p_1 + i\epsilon}
 \gamma^- \,
\int dx_2\,  \frac{(x_2-x_1) \gamma^+\gamma^-}{2x_2 + i\epsilon}
 \frac{\gamma^+}{2x_1 p_1 + i\epsilon} \hat\varepsilon^*  v(k)\times\,
\nonumber\\
&&
\Big\{
 (-)\int (d\lambda_1)\,
 e^{-ix_1\lambda_1}\,
\langle p_1, S^T| \bar\psi(0)\gamma^+ \psi(\lambda_1 n)\,
\int (d^4\ell) \delta(x_2-x_1-\ell n)\, a^+(\ell)| S^T, p_1\rangle
\Big\}\,,
\end{eqnarray}
where
\begin{eqnarray}
\label{B-par-QCD}
(-)\int (d\lambda_1)\,
 e^{-ix_1\lambda_1}\,
\langle p_1, S^T| \bar\psi(0)\gamma^+ \psi(\lambda_1 n)\,
\int (d^4\ell) \delta(x_2-x_1-\ell n) a^+(\ell)| S^T, p_1\rangle = B^V(x_1,x_2)\,.
\end{eqnarray}
As can be seen, however, that this diagram does not contribute to the Ward identity.
Indeed, after calculation of the imaginary
part we get the factor $(x_2-x_1)$ in the numerator of (\ref{QCD-WI-1}) which goes
to zero owing to $\delta(x_2-x_1)$ from $\Im\text{m}B^V(x_1,x_2)$.

Further, calculation of the H5-diagram, presented in Fig.\ref{Fig-All}, gives us
\begin{eqnarray}
\label{QCD-WI-2}
&&\overline{{\cal W}}(\text{diag.H5})=
\int dx_1\, dy \, \delta^{(4)}(x_1 p_1+y p_2-k-q) \,
{\cal F}^g(y)\, \text{C}_2 \times
\nonumber\\
&&\bar v(k)\hat\varepsilon \, \frac{\gamma^+}{2x_1 p_1 + i\epsilon}
 \gamma^- \,  \hat\varepsilon^*\,
\int dx_2\,  \frac{\gamma^+\gamma^-\gamma^+}{2x_2p_1 + i\epsilon}\, v(k)
\, B^V(x_1,x_2)\,,
\end{eqnarray}
while the contribution of the D1-diagram in Fig.\ref{Fig-All} takes the form
\begin{eqnarray}
\label{QCD-WI-3}
&&\overline{{\cal W}}(\text{diag.D1})=
- \int dx_1\, dy \, \delta^{(4)}(x_1 p_1+y p_2-k-q) \,
{\cal F}^g(y)\, \text{C}_1 \times
\nonumber\\
&&\bar v(k)\hat\varepsilon \, \frac{\gamma^+}{2x_1 p_1 + i\epsilon}
 \gamma^- \gamma^+ \gamma^-\,
\frac{\gamma^+}{2x_1p_1 + i\epsilon}\,  \hat\varepsilon^*\, v(k)
\int dx_2\, \, B^V(x_1,x_2)\,.
\end{eqnarray}
Finally, the contribution of the H9-diagram with the three-gluon vertex, see Fig.\ref{Fig-All},
reads
\begin{eqnarray}
\label{QCD-WI-4}
&&\overline{{\cal W}}(\text{diag.H9})=
-4\,i\, \int dx_1\, dy \, \delta^{(4)}(x_1 p_1+y p_2-k-q) \,
{\cal F}^g(y)\, \text{C}_3 \times
\nonumber\\
&&\bar v(k)\hat\varepsilon \, \frac{\gamma^+}{2x_1 p_1 + i\epsilon}
 \gamma^- \, \frac{\gamma^+}{2x_1p_1 + i\epsilon}\, \hat\varepsilon^*\, v(k)
\int dx_2\, \frac{x_2-x_1}{2(x_2-x_1)+i\epsilon} \, B^V(x_1,x_2)\,.
\end{eqnarray}
We now turn to the contour gauge. First of all, based on our previous discussions notice that even a fleeting
glance is sufficient to anticipate the corresponding prescriptions for $B^V$-functions in
(\ref{QCD-WI-2})--(\ref{QCD-WI-4}).
The H5-diagram in Fig.\ref{Fig-All} corresponds to the final state interaction
and, therefore, the function $B^V_-$
should appear here. On the other hand,  the D1- and H9-diagrams in Fig.\ref{Fig-All} correspond to the
initial state interaction which leads to the function $B^V_+$.
Performing the explicit calculations (see also \cite{AT-10}),
we can arrive at the conclusion mentioned above by restoring the Wilson lines in the quark-gluon correlators
of the mentioned diagrams.
The Wilson line, $[+\infty^-,\, z^-]$,  enters in the hadron tensor represented by
the H5-diagram in Fig.\ref{Fig-All} while
the Wilson line, $[z^-,\,-\infty^- ]$, appears in the hadron tensor represented by
the D1- and H9-diagrams in Fig.\ref{Fig-All}.

We sum all contributions and get the following final expression:
\begin{eqnarray}
\label{QCD-WI-final}
\sum\limits_{\text{\tiny N=H1,H5,D1,H9}}\overline{{\cal W}}(\text{diag. N})&=&
\frac{\text{C}_2}{8 x_1} \gamma^+ \gamma^- \gamma^+ \gamma^- \gamma^+
\int dx_2 \frac{B^V_-(x_1,x_2)}{x_2} +
\frac{\text{C}_1}{8 x_1^2} \gamma^+ \gamma^- \gamma^+ \gamma^- \gamma^+
\int dx_2 B^V_+(x_1,x_2) +
\nonumber\\
&&
\frac{i\text{C}_3}{4 x_1^2} \gamma^+ \gamma^- \gamma^+
\int dx_2 \frac{(x_2-x_1) B^V_+(x_1,x_2)}{x_2-x_1+i\epsilon}\,
\end{eqnarray}
where the function $B^V_{-}$ are represented by (\ref{B-minus}) or (\ref{B-fsi}), and
the function $B^V_+$ is given by (\ref{B-plus}) or
\begin{eqnarray}
\label{B-isi}
B^V_{+}(x_1,x_2)\equiv B^V_{\text{ISI}}(x_1,x_2)= \frac{T(x_1,x_2)}{x_1 - x_2 + i\epsilon}\,.
\end{eqnarray}
We are now calculating the imaginary part and, ultimately, we derive the QCD Ward identity in the form
\begin{eqnarray}
\label{QCD-WI}
\text{C}_2 - \text{C}_1 - i\text{C}_3 =
- [t^a, t^b]\, t^b\, t^a + i f^{abc} t^c\, t^b\, t^a  \equiv 0\,.
\end{eqnarray}
We want to stress that the identity (\ref{QCD-WI}) takes place provided only the
presence of the different complex prescriptions in gluonic poles dictated by the final or
initial state interactions:
\begin{eqnarray}
\label{QCD-illust}
  \begin{aligned}
\text{\bf FSI}\Rightarrow \frac{1}{\ell^+ + i\epsilon} \Rightarrow \text{gauge}\,\,\,[+\infty^-,\, z^- ]=1
\Rightarrow \frac{T(x_1,x_2)}{x_1-x_2 - i\epsilon}\,\\
\text{\bf ISI}\Rightarrow \frac{1}{-\ell^+ + i\epsilon} \Rightarrow \text{gauge}\,\,\,[z^-,\, -\infty^-]=1
\Rightarrow \frac{T(x_1,x_2)}{x_1-x_2 +i\epsilon}\,\\
  \end{aligned}\,\,\, \Bigg\}
\Rightarrow\text{\bf QCD GI}\,.
\end{eqnarray}
We emphasize on the principle differences (see details in Appendix~\ref{App:B}) between the considered case and the proof
of the QCD gauge invariance for the perturbative Compton scattering amplitude with the physical gluons in the
initial and final states. The latter does not need any external condition such as the presence of gluon poles.

Thus, the situation which we discuss is  again absolutely similar to that one which has been described in
\cite{Braun} for the dijet production. From (\ref{QCD-illust}), it is seen that
the different diagrams correspond
to the different contour gauges and, consequently, to the different
functions, $B^V_{\pm}$, which parametrize the hadronic
matrix element forming the soft part. In this context, we also have
a soft breaking of factorization because, first, it spoils
the universality principle and, second, the gluonic pole prescriptions in
the soft part are traced to the causal prescriptions in
the hard part. At the same time, we can use the replacement \cite{Braun}:
\begin{eqnarray}
\label{rep}
\frac{T(x_1,x_2)}{x_1-x_2 - i\epsilon} =
\frac{T(x_1,x_2)}{x_1-x_2 + i\epsilon} + 2\pi\, i\, \delta(x_1-x_2)T(x_1,x_2)
\end{eqnarray}
and finally get the same function $B^V_+$ for all diagrams (in other words, we can use the same contour gauge
for all diagrams). However, it contains the additional $\delta(x_1-x_2)$-term which may lead
to the collinear factorization
violation in the same manner as in \cite{Braun}.
The full analysis of this case will be implemented in our forthcoming work.

\section{Hadron tensor of the direct photon production II:
new contributions}

\begin{figure}[t]
\centerline{\includegraphics[width=0.5\textwidth]{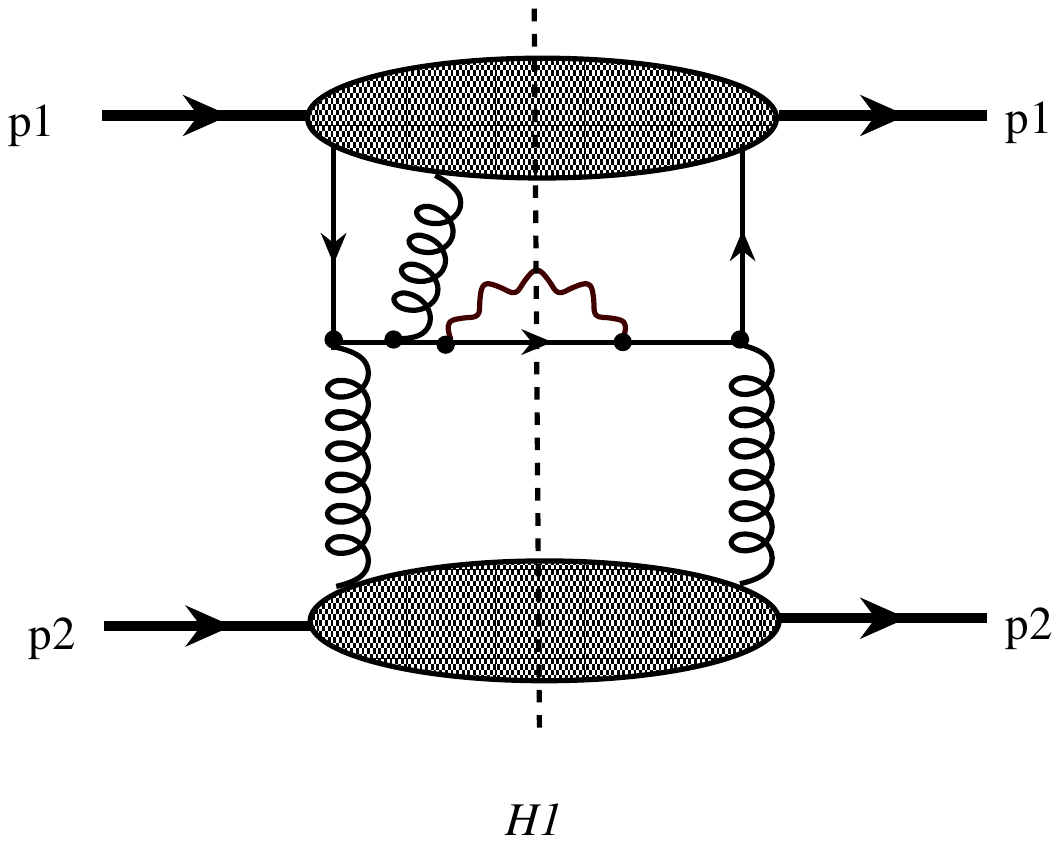}
\hspace{-4.cm}\includegraphics[width=0.5\textwidth]{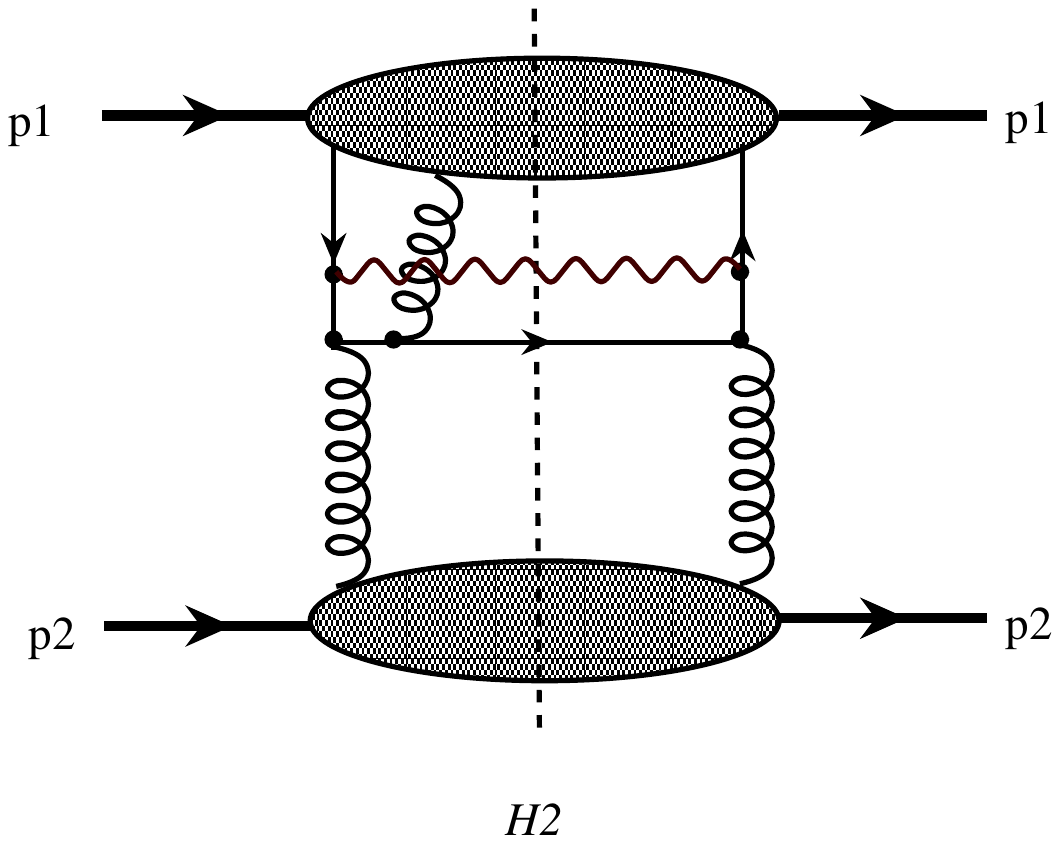}
\hspace{-4.cm}\includegraphics[width=0.5\textwidth]{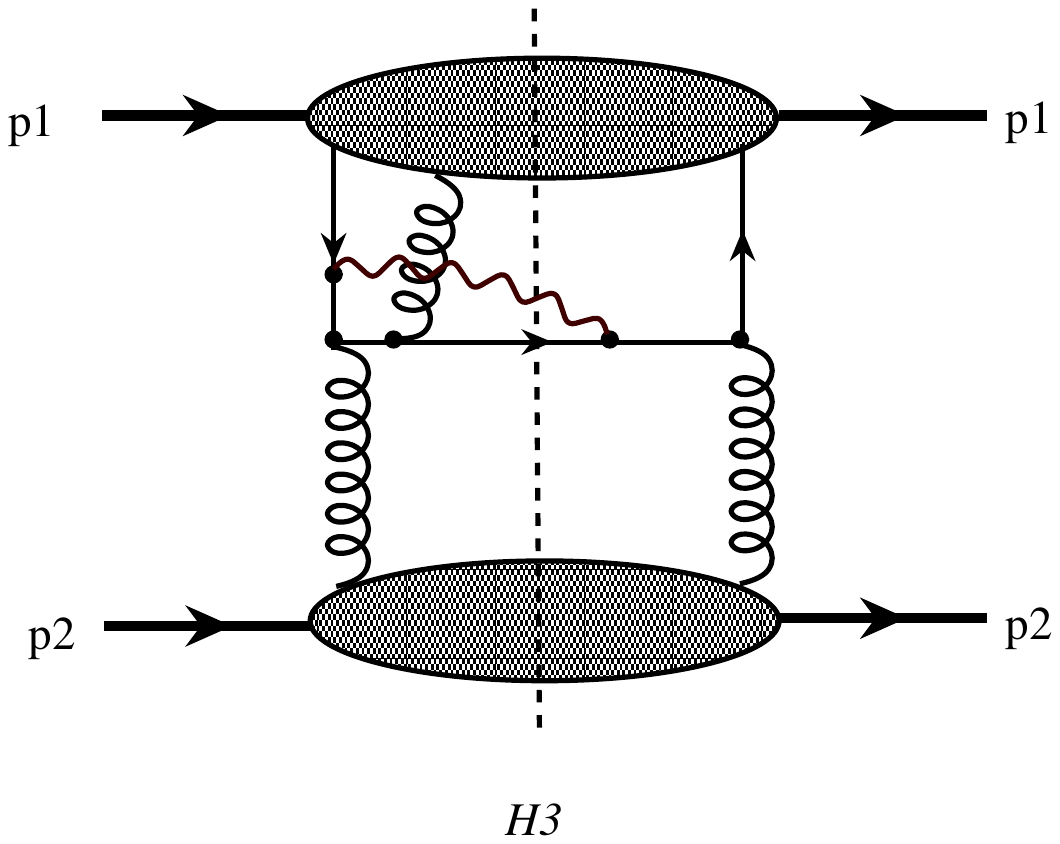}
}
\vspace{-8.5cm}
\centerline{\includegraphics[width=0.5\textwidth]{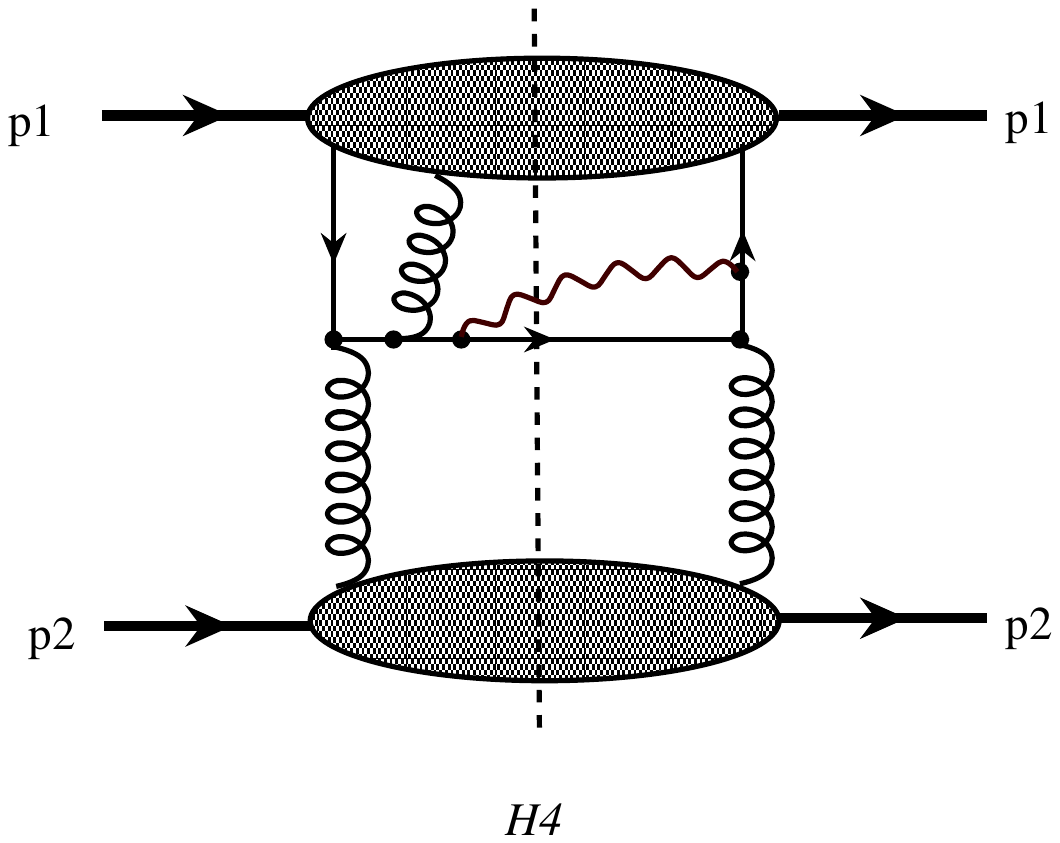}
\hspace{-4.cm}\includegraphics[width=0.5\textwidth]{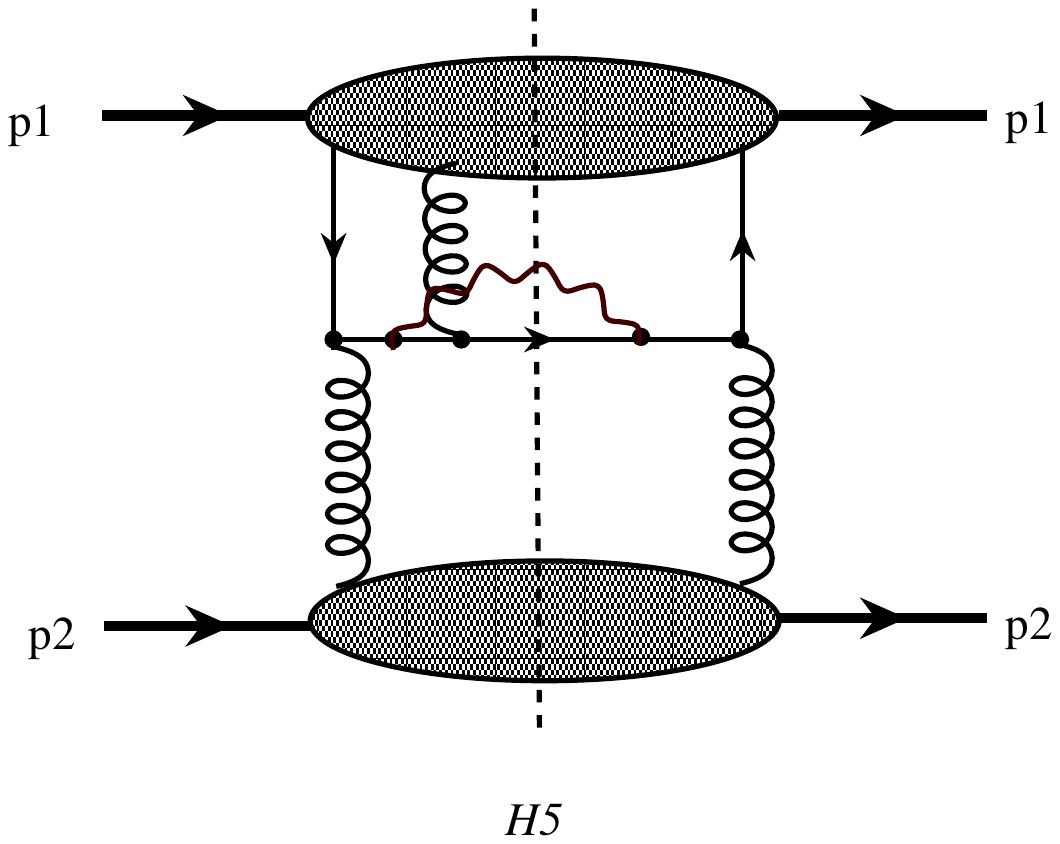}
\hspace{-4.cm}\includegraphics[width=0.5\textwidth]{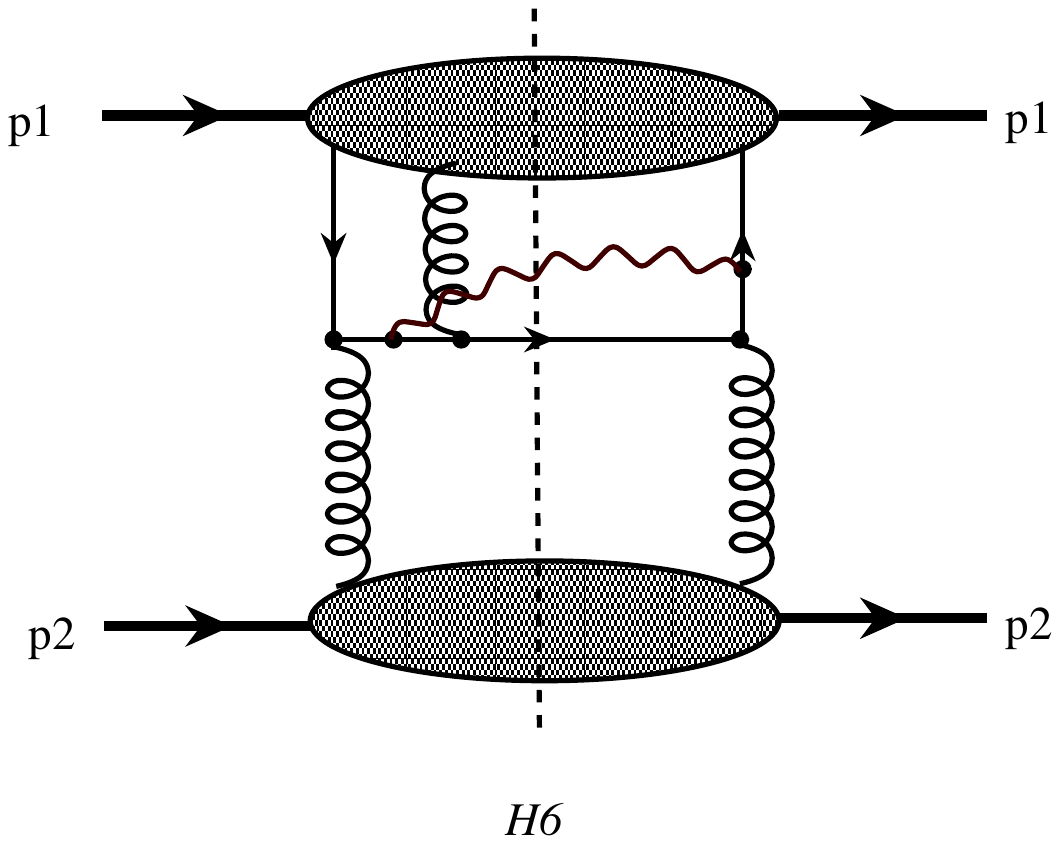}
}
\vspace{-8.5cm}
\centerline{\includegraphics[width=0.5\textwidth]{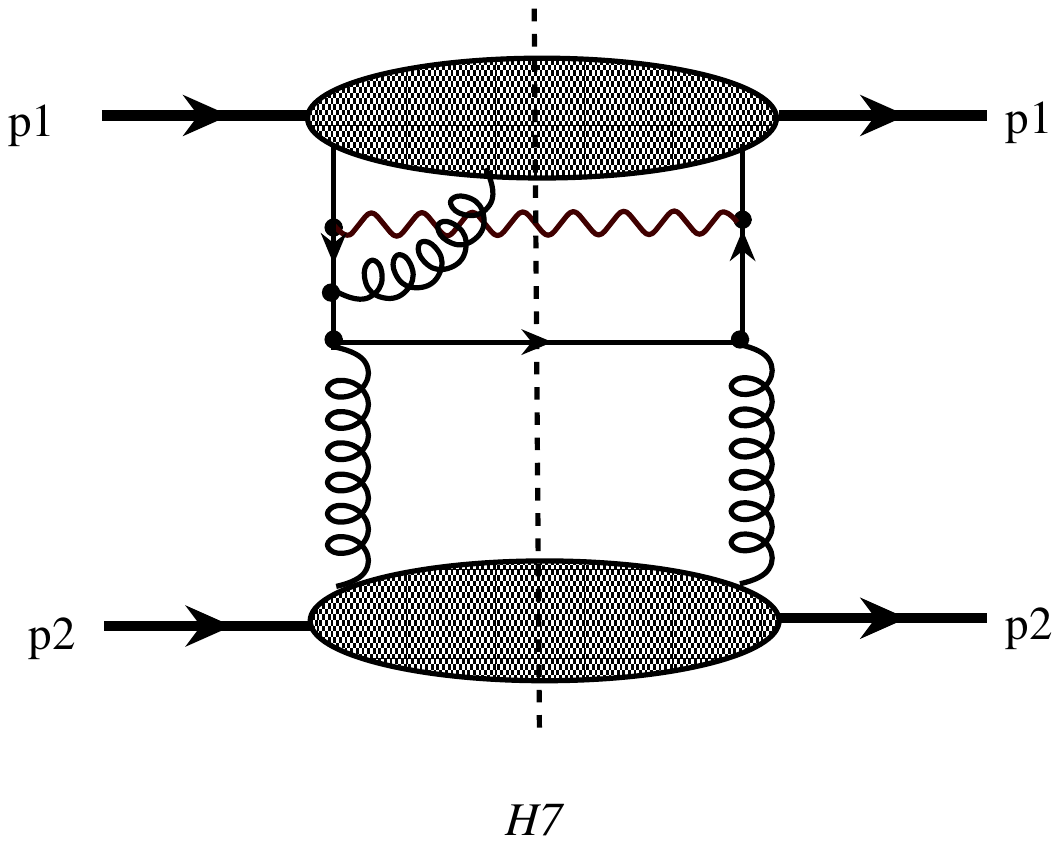}
\hspace{-4.cm}\includegraphics[width=0.5\textwidth]{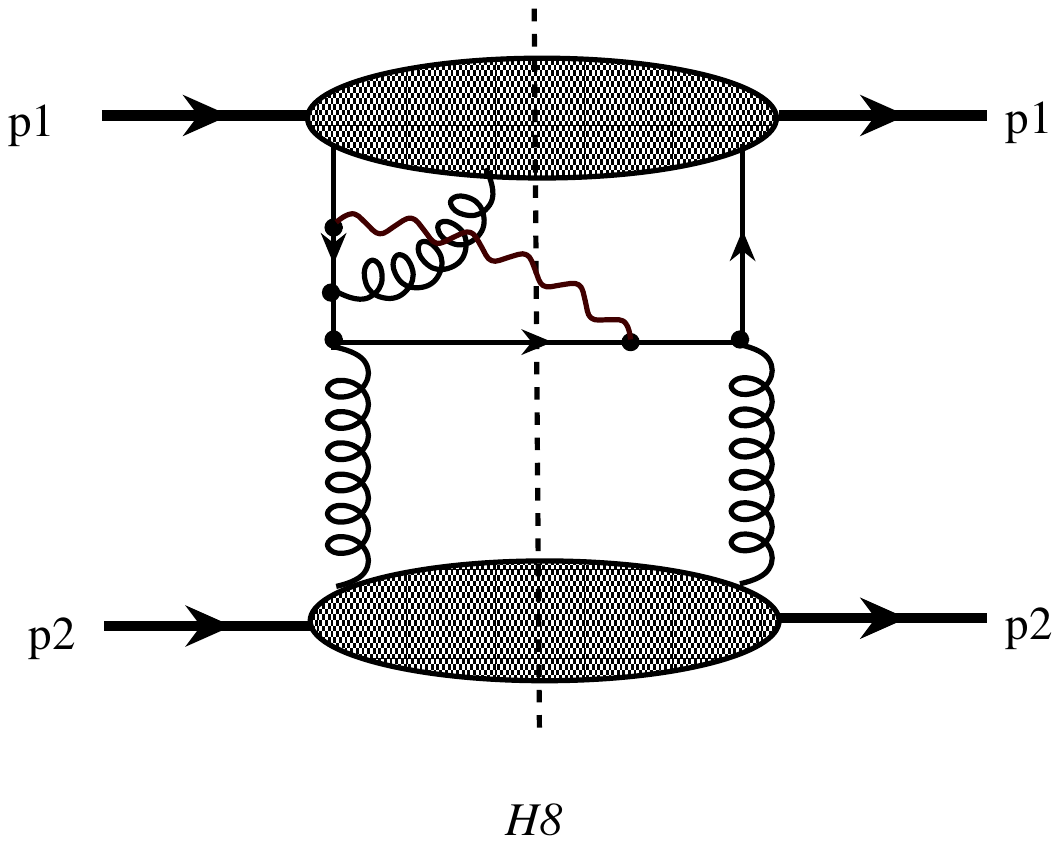}
\hspace{-4.cm}\includegraphics[width=0.5\textwidth]{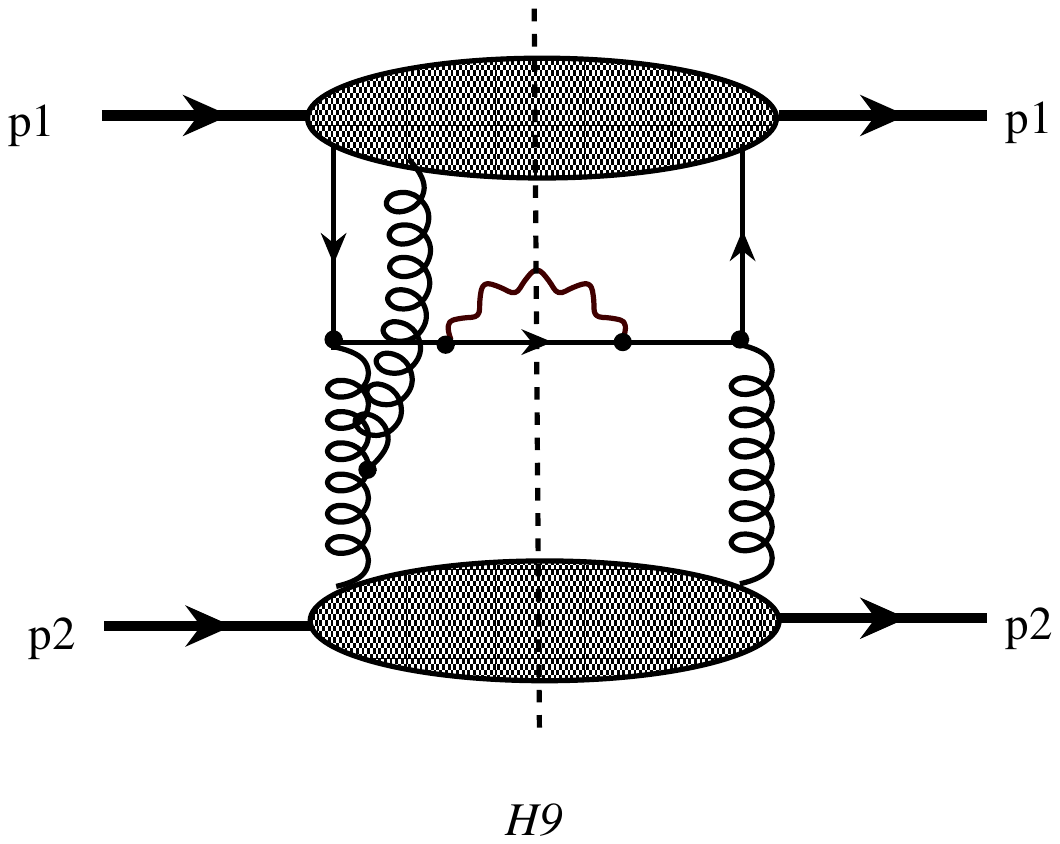}
}
\vspace{-8.5cm}
\centerline{\includegraphics[width=0.5\textwidth]{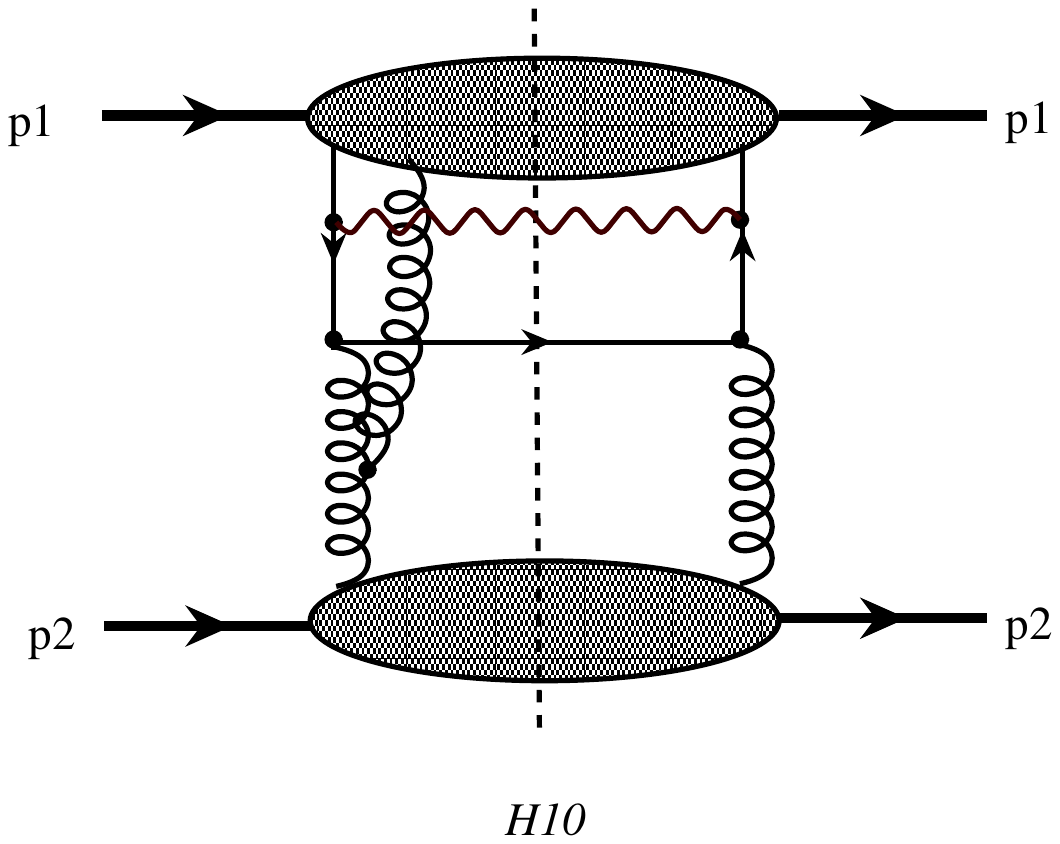}
\hspace{-4.cm}\includegraphics[width=0.5\textwidth]{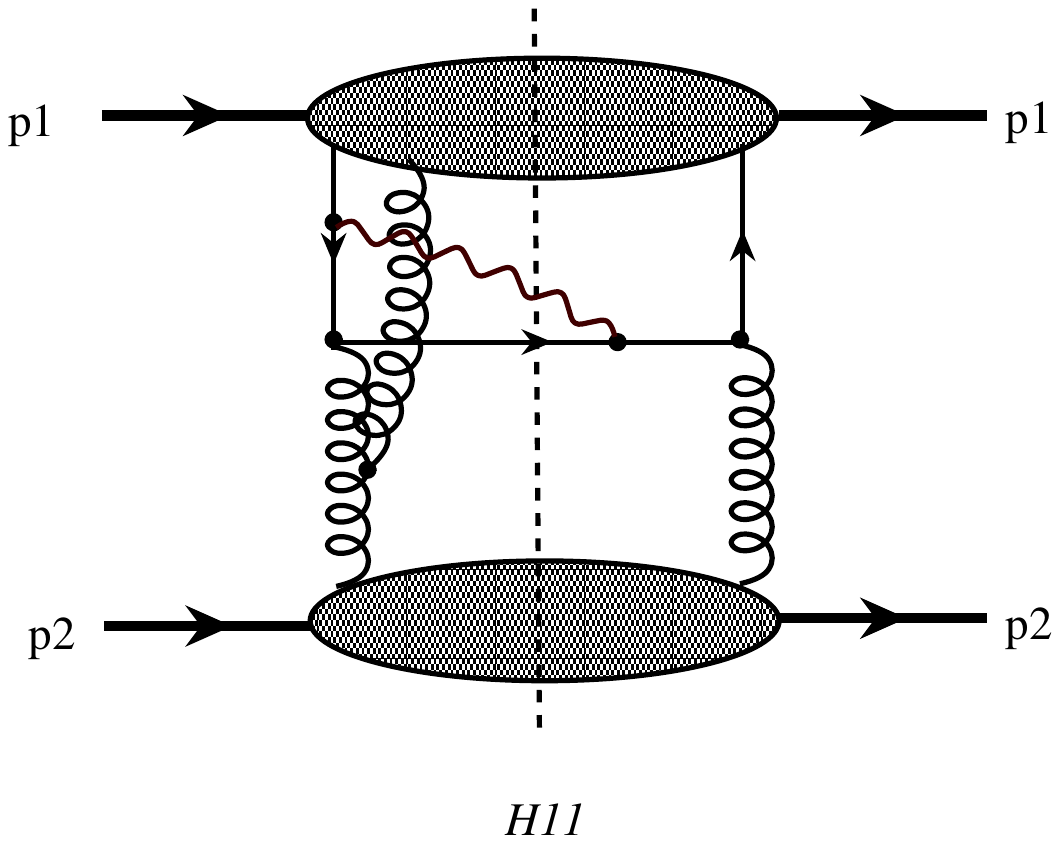}
\hspace{-4.cm}\includegraphics[width=0.5\textwidth]{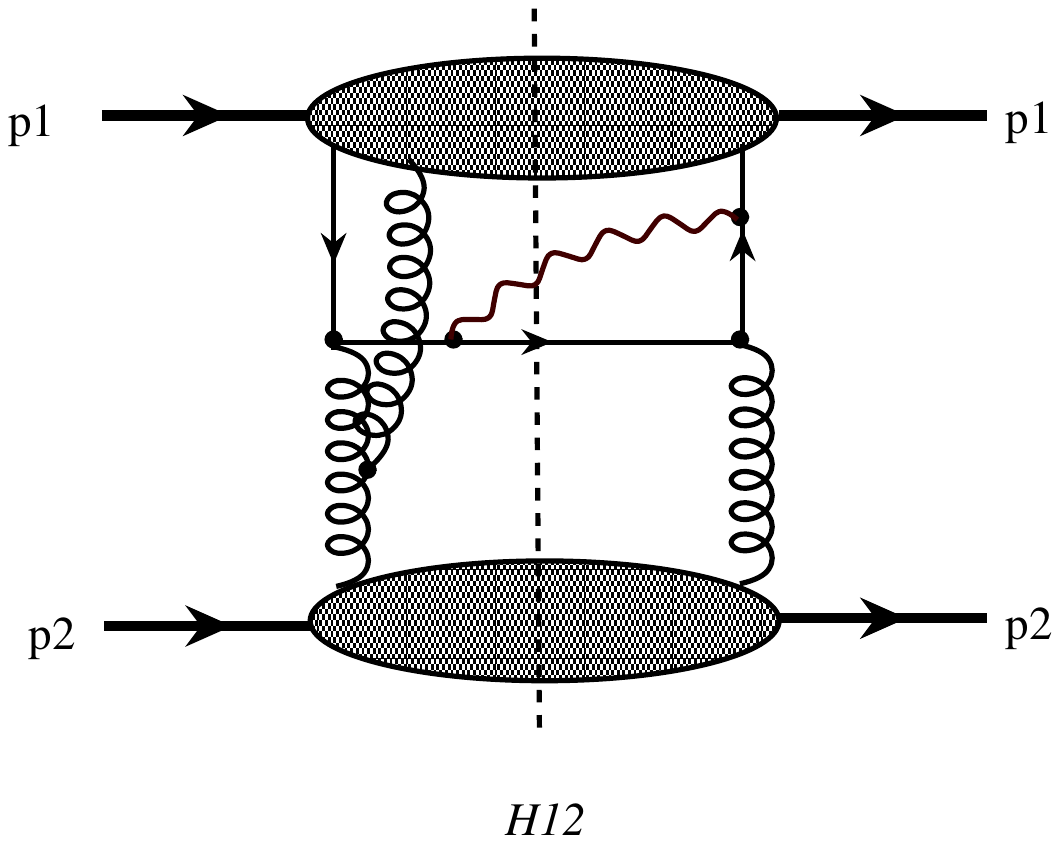}
}
\vspace{-8.5cm}
\centerline{\includegraphics[width=0.5\textwidth]{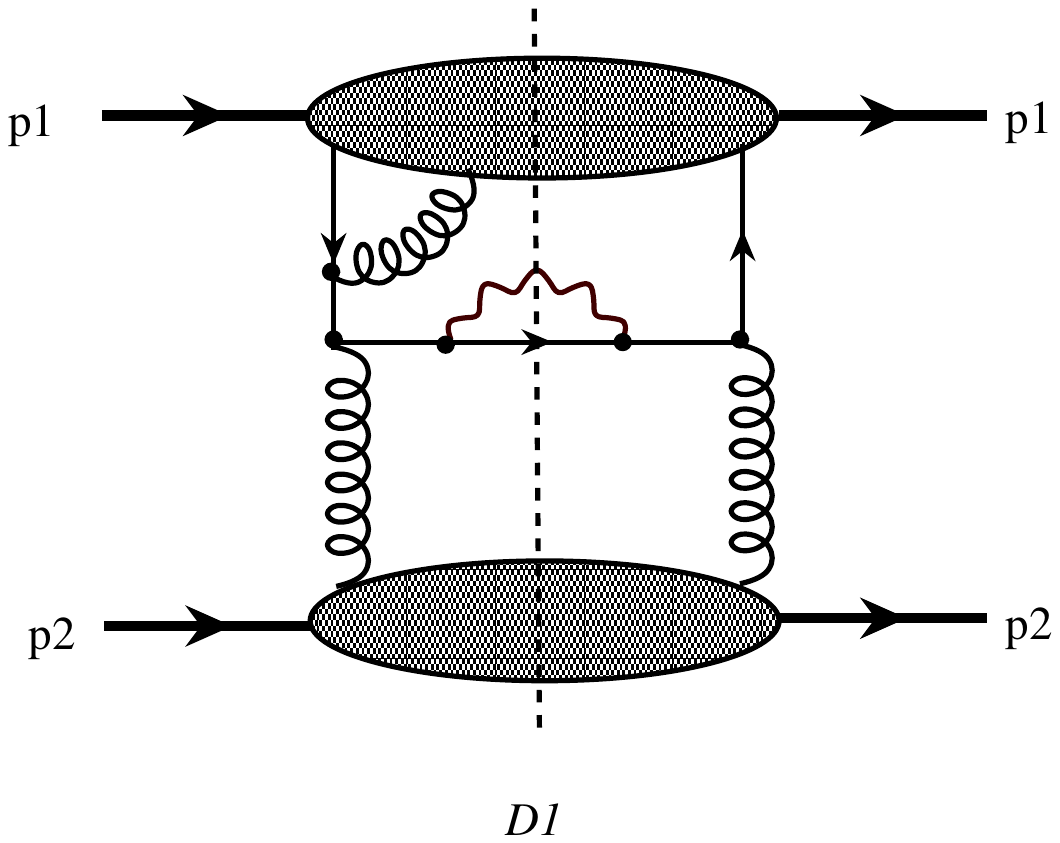}
\hspace{-4.cm}\includegraphics[width=0.5\textwidth]{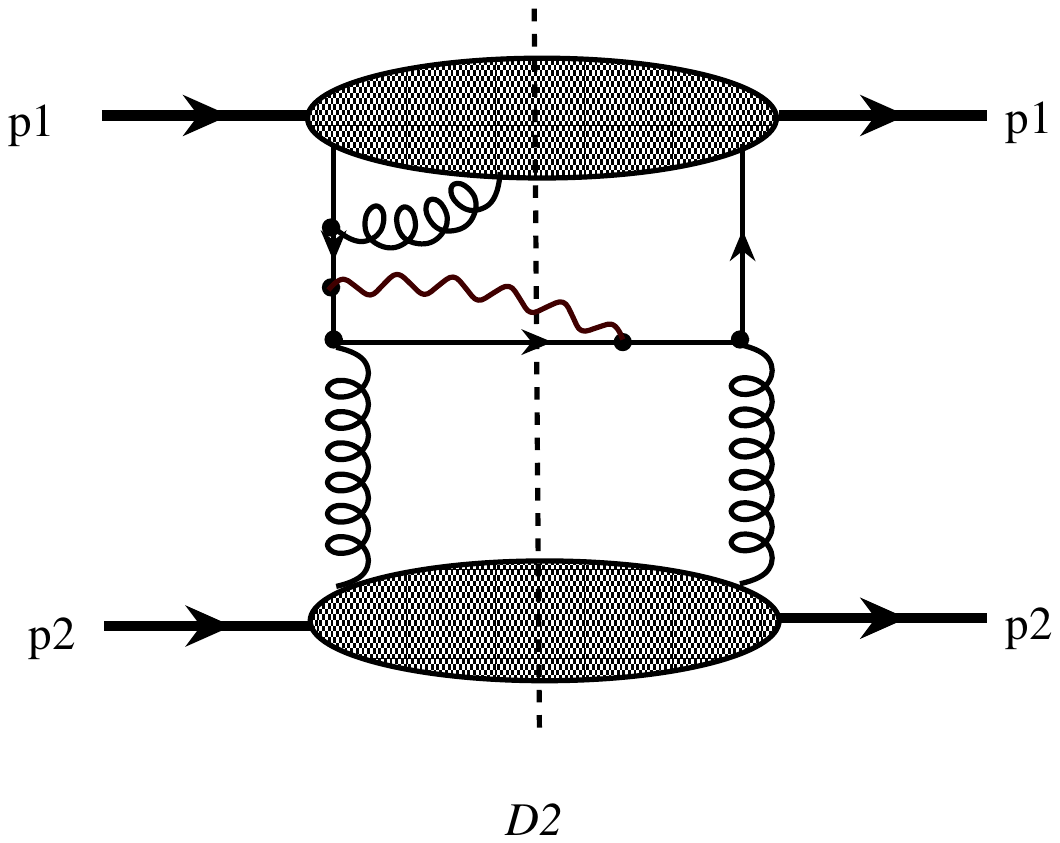}
\hspace{-4.cm}\includegraphics[width=0.5\textwidth]{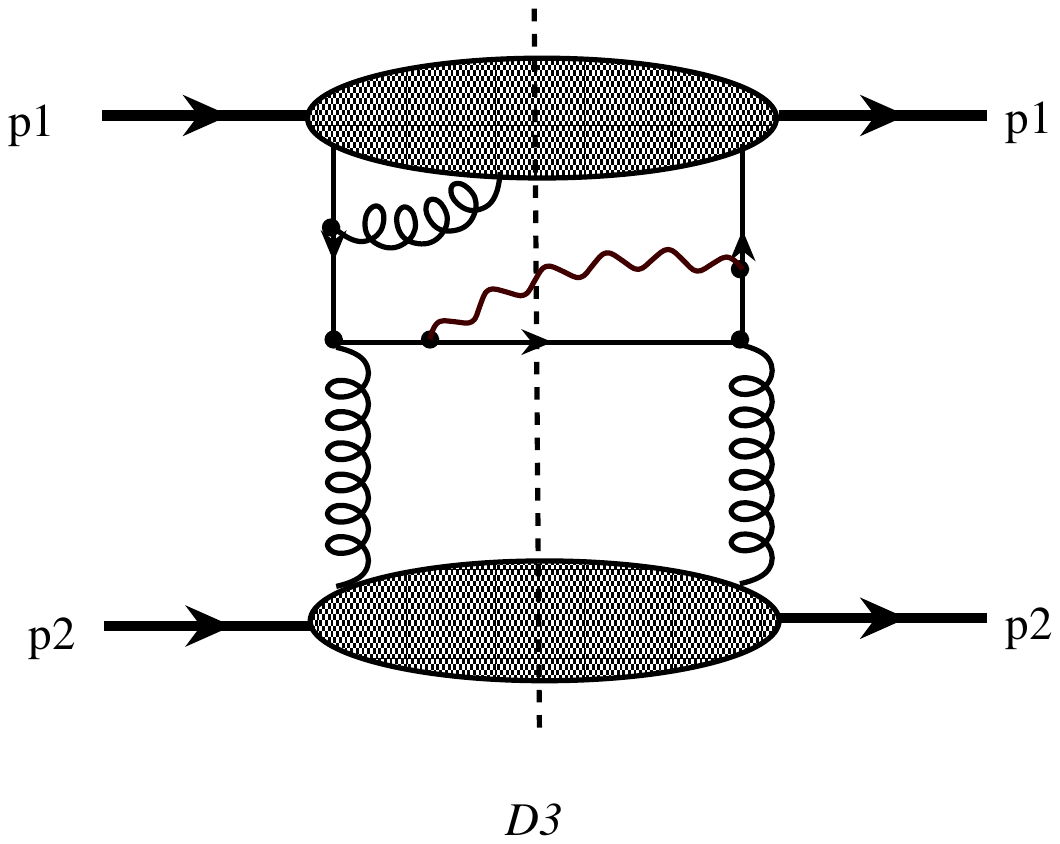}
}
\vspace{-8.5cm}
\centerline{\includegraphics[width=0.5\textwidth]{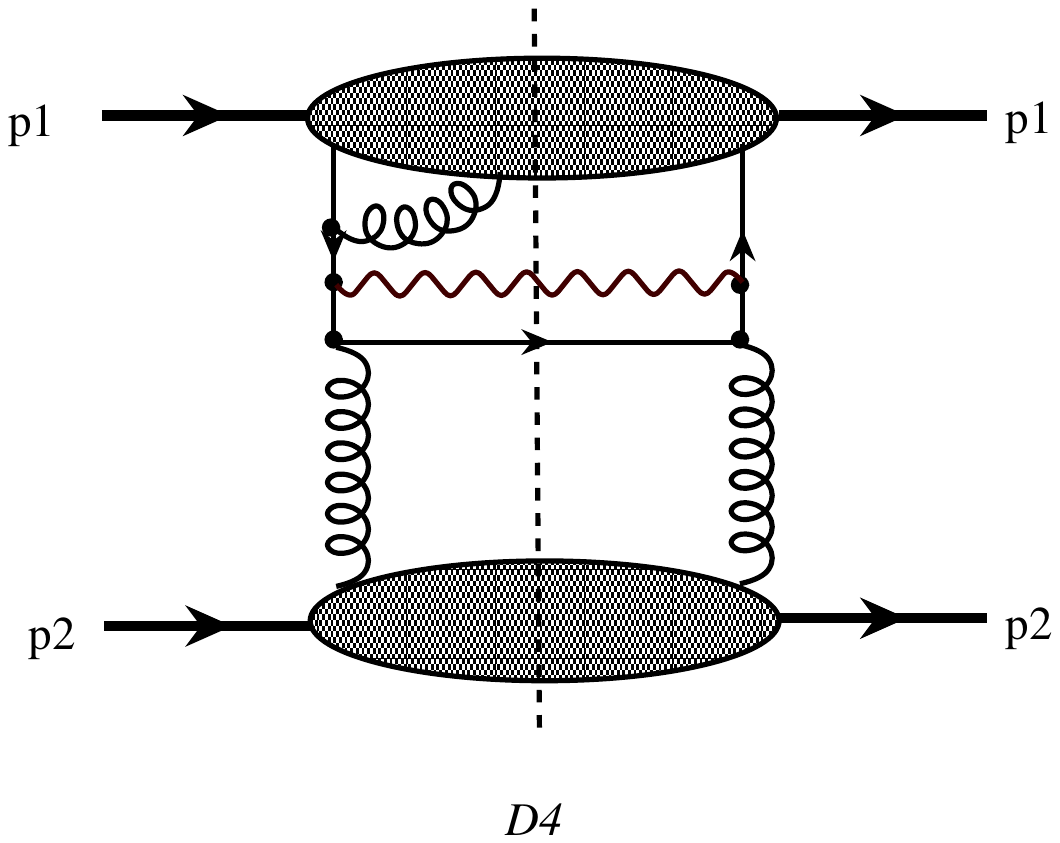}
}
\vspace{-7.cm}
\caption{The typical Feynman diagrams, contributing to the hadron tensor.}
\label{Fig-All}
\end{figure}

In this section we calculate the full expression for the hadron tensor
which involves both the standard and new contributions to the gluon pole terms.
The full expression for the hadron tensor related
to the case we are discussing can be split into two groups:
(i) the first type of contributions corresponds to the diagrams H1--H12 depicted in Fig.~\ref{Fig-All} and, before factorization,
takes the following form
\begin{eqnarray}
\label{GroupI}
{\cal W}(\text{diag.H})&=&\int\frac{d^3 \vec{q}}{(2\pi)^3 2 E} \frac{d^3 \vec{k}}{(2\pi)^3 2 \varepsilon}
\, \text{C}_H\, \int (d^4 k_1) (d^4 k_2) \delta^{(4)}(k_1+k_2-q-k)\,  \Phi_g^{\alpha\beta}(k_2)\, \times
\nonumber\\
&&\int (d^4\ell)\,
\Phi^{[\gamma^+],\,\rho}_\perp(k_1,\ell)\, H^{\alpha\beta, \rho}(k_1,k_2,\ell) \,,
\end{eqnarray}
and (ii) the second type of contributions is given by the diagrams D1--D4 in Fig.~\ref{Fig-All} can be presented as
\begin{eqnarray}
\label{GroupII}
{\cal W}(\text{diag.D})&=& \int\frac{d^3 \vec{q}}{(2\pi)^3 2 E} \frac{d^3 \vec{k}}{(2\pi)^3 2 \varepsilon}
\, \text{C}_D\, \int (d^4 k_1) (d^4 k_2) \delta^{(4)}(k_1+k_2-q-k)\,\times
\nonumber\\
&&\Phi_g^{\alpha\beta}(k_2)\,
{\rm tr}_{\it D}\big[ \Phi^{(1)}(k_1)\, D^{\alpha\beta}(k_1,k_2)\big] \,.
\end{eqnarray}
In eqns.~(\ref{GroupI}) and (\ref{GroupII}), the corresponding coefficient functions are denoted by $H_{\alpha\beta, \rho}(k_1,k_2,\ell)$ and
$D_{\alpha\beta}(k_1,k_2)$. The unpolarized twist-$2$ gluon distribution $\Phi^g(k_2)$ and the twist-$3$ quark distribution
$\Phi_\perp^{[\gamma^+],\rho}(k_1,\ell)$ are defined in the standard forms, see (\ref{g-func}) and (\ref{q-func}).
The twist-$3$ quark distribution which appears in the diagrams D1--D4 presented in Fig.~\ref{Fig-All} is given by
\begin{eqnarray}
\label{q-func-2}
\Phi^{(1)}(k_1)=  \frac{\gamma^+ \gamma^\rho_\perp\, \gamma^-}{2k_1^++i\epsilon}\,
\int (d^4\eta_1)\,
 e^{ik_1\eta_1}\,
\langle p_1, S^T| \bar\psi(0)\gamma^+ A^\rho_\perp(0)\psi(\eta_1) | S^T, p_1\rangle\,,
\end{eqnarray}
where the sum over the corresponding intermediate states is implied.
We now perform the factorization procedure for Eqns. (\ref{GroupI}) and (\ref{GroupII}), and  we
obtain
\begin{eqnarray}
\label{GroupI-f}
d{\cal W}(\text{diag.H})&=&\frac{d^3 \vec{q}}{(2\pi)^3 2 E} \int \frac{d^3 \vec{k}}{(2\pi)^3 2 \varepsilon}
\delta^{(2)}(\vec{\bf k}_\perp+\vec{\bf q}_\perp)
\, \text{C}_H\, \int dx_1 dy \delta(x_1-x_B)\,\delta(y-y_B)\,\frac{2}{S}  {\cal F}^g(y)\,g_\perp^{\alpha\beta}\, \times
\nonumber\\
&&\int dx_2\,
\Phi^{[\gamma^+],\,\rho}_\perp(x_1,x_2)\, H^{\alpha\beta, \rho}(x_1,x_2) \,,
\end{eqnarray}
for the first type of contributions and
\begin{eqnarray}
\label{GroupII-f}
d{\cal W}(\text{diag.D})&=&\frac{d^3 \vec{q}}{(2\pi)^3 2 E} \int \frac{d^3 \vec{k}}{(2\pi)^3 2 \varepsilon}
\delta^{(2)}(\vec{\bf k}_\perp+\vec{\bf q}_\perp)
\, \text{C}_D\, \times
\nonumber\\
&&\int dx_1 dy \delta(x_1-x_B)\,\delta(y-y_B)\,\frac{2}{S}  {\cal F}^g(y)\,g_\perp^{\alpha\beta}\,
{\rm tr}_{\it D}\big[\Phi^{(1)}(x_1)\, D^{\alpha\beta}(x_1)\big] \,,
\end{eqnarray}
for the second type of contributions.

To simplify our calculations without losing generality, we may impose the frame
where $q_\perp^2\ll S$. The Mandelstam variable defined for the subprocess, $\hat u$, is a small
variable and can be neglected. It means that the Bjorken fraction $y_B$ becomes independent of $x_B$,
and one can write $y_B=-T/S$ (due to $\hat s + \hat t + \hat u =0$).

The next stage is to determine the type of the twist-$3$ function $B^V(x_1,x_2)$ which is related to the certain
complex prescriptions in gluon poles according to the way described in the preceding sections.
The correct definition of $B^V(x_1,x_2)$-function should be implemented for each of diagrams.
Also, it is instructive to notice that
the diagrams H1--H8 would not possess the gluon poles
in the case where the function $B^V$ assumed to be a real one. However, this is not true in our case.

After computing the corresponding traces and performing simple algebra within the frame we are choosing,
it turns out that the only nonzero contributions to the hadron tensor come from the diagrams H1, H7, D4 and H10:
\begin{eqnarray}
\label{diaH1}
d{\cal W}(\text{diag.H1})&=&\frac{d^3 \vec{q}}{(2\pi)^3 2 E} \int \frac{d^3 \vec{k}}{(2\pi)^3 2 \varepsilon}
\delta^{(2)}(\vec{\bf k}_\perp+\vec{\bf q}_\perp)
\, \text{C}_2\, \int dx_1 dy \delta(x_1-x_B)\,\delta(y-y_B)\,  {\cal F}^g(y)\, \times
\nonumber\\
&&\int dx_2\, \frac{2 S^2\, x_1 \, y^2}{[x_2 y S + i\epsilon][x_1 y S + i\epsilon]^2}\,
\frac{\varepsilon^{q_\perp + S_\perp -}}{p_1^+}\, B^{V}_-(x_1,x_2)\,,
\end{eqnarray}
\begin{eqnarray}
\label{diaH7}
d{\cal W}(\text{diag.H7})&=&\frac{d^3 \vec{q}}{(2\pi)^3 2 E} \int \frac{d^3 \vec{k}}{(2\pi)^3 2 \varepsilon}
\delta^{(2)}(\vec{\bf k}_\perp+\vec{\bf q}_\perp)
\, \text{C}_1\, \int dx_1 dy \delta(x_1-x_B)\,\delta(y-y_B)\, {\cal F}^g(y)\, \times
\nonumber\\
&&\int dx_2\, \frac{(-2) S\, T\, x_1\, (y-3y_B)}{[x_2 T + i\epsilon][x_1 T + i\epsilon]^2}\,
\frac{\varepsilon^{q_\perp + S_\perp -}}{p_1^+}\, B^{V}_+(x_1,x_2)\,,
\end{eqnarray}
\begin{eqnarray}
\label{diaD4}
d{\cal W}(\text{diag.D4})&=&\frac{d^3 \vec{q}}{(2\pi)^3 2 E} \int \frac{d^3 \vec{k}}{(2\pi)^3 2 \varepsilon}
\delta^{(1)}(\vec{\bf k}_\perp+\vec{\bf q}_\perp)
\, \text{C}_1\, \int dx_1 dy \delta(x_1-x_B)\,\delta(y-y_B)\,\frac{2}{S}  {\cal F}^g(y)\, \times
\nonumber\\
&&\frac{2 S^2 \,x_1\, (y-2y_B)}{[x_1 T + i\epsilon]^2}\,
\frac{\varepsilon^{q_\perp + S_\perp -}}{2x_1p_1^+ + i\epsilon}\, \int dx_2\,B^{V}_+(x_1,x_2)\,,
\end{eqnarray}
and
\begin{eqnarray}
\label{diaH10}
d{\cal W}(\text{diag.H10})&=&\frac{d^3 \vec{q}}{(2\pi)^3 2 E} \int \frac{d^3 \vec{k}}{(2\pi)^3 2 \varepsilon}
\delta^{(2)}(\vec{\bf k}_\perp+\vec{\bf q}_\perp)
\,\text{C}_3\, \int dx_1 dy \delta(x_1-x_B)\,\delta(y-y_B)\, {\cal F}^g(y)\, \times
\nonumber\\
&&\int dx_2\, \frac{2 T (x_1-x_2) (2 T+S y)}{[x_1 T + i\epsilon][x_2 T + i\epsilon][(x_1-x_2) y S + i\epsilon]}\,
\frac{\varepsilon^{q_\perp + S_\perp -}}{p_1^+}\, B^{V}_+(x_1,x_2)\,.
\end{eqnarray}
Here, $\text{C}_1=C_F^2 N_c$, $\text{C}_2=-C_F/2$, $\text{C}_3=C_F\, N_c \,C_A/2$.
The other diagram contributions disappear owing to the following reasons: (i) the $\gamma$-algebra gives
$(\gamma^-)^2=0$; (ii) the common pre-factor $T+yS$ goes to zero,
(iii) the diagrams H2 and H5 cancel each other.

Analysing the results for the diagrams  H1, H7, D4 and H10 (see Eqns. (\ref{diaH1})--(\ref{diaH10})),
we can see that
\begin{eqnarray}
\label{Fac2}
d{\cal W}(\text{diag.H1}) + d{\cal W}(\text{diag.H7}) + d{\cal W}(\text{diag.D4}) =
d{\cal W}(\text{diag.H10})\,.
\end{eqnarray}
In other words, as similar to the Drell-Yan process, the new (``non-standard") contributions
generated by the diagrams H1, H7 and D4 result again in the factor of $2$ compared to the
``standard" diagram H10 contribution to the corresponding hadron tensor.
This is our principle result.

%%%%%%%%%%%%%%%%%%%%%%%%%%%%% FIGURE %%%%%%%%%%%%%%%%%%%%%%%%%%%%%%%%
%%%%%%%%%%%%%%%%%%%%%%%%%%%%%%%%%%%%%%%%%%%%%%%%%%%%%%%%%%%%%%%%%%%%%%%

\section{Conclusions}

In this work, we explore both the QED and QCD gauge invariance of
the hadron tensor for the direct photon production in two
hadron collision where one of hadrons is transversely polarized.
We present the important details related to the use of
the contour gauge for the Drell-Yan process with essential
transverse polarizations.
We study the effects which lead to the soft breaking of factorization
through the QED and QCD gauge invariance.

We show that the contour gauges for gluon fields play
the crucial role for our study. The contour gauge belongs to the class of non-local  gauges
that depends on the path connecting two points in the correlators.
It turns out, in the cases which we consider, the prescriptions for the gluonic poles in the twist $3$ correlators
are dictated by the prescriptions in the corresponding hard parts in the similar manner as for the DY-process
considered in \cite{AT-10}.

For the direct photon production, we demonstrate that the prescriptions
in the gluonic pole contributions differ from each other
depending upon the initial or final state interactions in the related diagrams.
We stress that the different prescriptions are needed to ensure the QCD gauge invariance.
This situation has been treated as a soft breaking of the
universality condition resulting in factorization breaking.
Besides, the presence of the complex prescriptions in the gluonic pole contributions allow
the extra diagrams to contribute nontrivially to the hadron tensor.

We find that the ``non-standard" new terms,
which exist in the case of the complex twist-$3$ $B^V$-function with the corresponding prescriptions,
do contribute to the hadron tensor exactly as the ``standard" term known previously.
This is  another important result of our work.
We also observe that this is exactly similar to the case of Drell-Yan process studied in \cite{AT-10}.

\section{Acknowledgements}

We would like to thank M.~Deka, I.O.~Cherednikov, A.V.~Efremov, D.~Ivanov and
L.~Szymanowski for useful discussions and correspondence.
This work is partly supported by the HL program.

%%%%%%%%%%%%%%%%%%%%%%%%%%%%%%%%%%%%%%%%%%%%%%%%%%%%%%%%%%%%%%%%%%%%%%%%%%%%%
%%%%%%%%%%%%%%%%%%%%%%%%%%%%%   Appendix   %%%%%%%%%%%%%%%%%%%%%%%%%%%%%%%%%%
%%%%%%%%%%%%%%%%%%%%%%%%%%%%%%%%%%%%%%%%%%%%%%%%%%%%%%%%%%%%%%%%%%%%%%%%%%%%%
\appendix
\renewcommand{\theequation}{\Alph{section}.\arabic{equation}}
\section*{Appendix}
%%%%%%%%%%%%%%%%%%%%%%%%%%%%%%%%%%%%%%%%%%%%%%%%%%%%%%%%%%%%%%%%%%%%%%%%%%%%%%%%%%%%%%%%%%

\section{$B^V(x_1,x_2)-$function within the contour gauge}
\label{ContourGauge:App:A}

In this Appendix we give some details regarding the use of the contour gauge.
The solution of the QED gauge invariance problem
for the DY hadron tensor can be found by using the contour
gauge conception (see, for example, \cite{ContourG, AT-10}).
Here, we would  like to demonstrate that there is no an ambiguity in the integral representation
 of the gluon fields through the strength tensor
(Eqns. (\ref{Ag1}) and (\ref{Ag2})).
It is important to note that the axial gauge condition, $A^+=0$,
(as well as the Fock-Schwinger gauges)
is actually a particular case of the most general contour gauge
where the Wilson line with an arbitrary path determines the gauge transformations.
The contour gauge was the subject of very intense studies
many years ago. The preponderances of the use of the contour gauge
is that the quantum gauge theory becomes  free from the Gribov ambiguities.

Let us briefly discuss the main items of the contour gauge conception.
To describe this class of gauges, it is instructive to assume the geometrical interpretation of gluons
where the gluon field is a connection of the principal fiber bundle
${\cal P}(\mathbb{R}^4, G, \pi)$
(here, $\mathbb{R}^4$ implies the base where the principal fiber bundle is determined, $G$ denotes the group
defined on the given fiber and $\pi$ is a transformation of the base $\mathbb{R}^4$ into the fiber bundle
${\cal P}$). Each element $\textbf{g}(x)$ of the fiber, with the help of the gluon field
$A_\alpha$, defines the gauge-transformed field:
\begin{eqnarray}
\label{Ag}
A^{\textbf{g}}_\mu(x)=\textbf{g}^{-1}(x)A_{\mu}(x) \textbf{g}(x) +
\frac{i}{g}\textbf{g}^{-1}(x)\partial_{\mu}\textbf{g}(x).
\end{eqnarray}
The set of these fields for all $\textbf{g}(x)$ forms the orbit of the gauge-equivalent fields.
It is  well-known that in order to quantize the system of the gauge fields, one has to choose the only element of each
orbit. In contrast to the usual way, we first fix an arbitrary
point $(x_0, \textbf{g}(x_0))$ \footnote{We assume that the subspace in the surroundings of an arbitrary
point belonging to the fiber can be trivialized. That means we can introduce the co-ordinate of the point
$(x,\textbf{g}(x))$.} in the fiber. Then, we define two directions: one of them in the base, the other in
the fiber. The direction in the base $\mathbb{R}^4$ is nothing else than the tangent vector of a curve which
goes through the given point $x_0$. At the same time, the direction in the fiber can be uniquely determined as the
tangent subspace which is related to the parallel transition. After following this procedure, one can uniquely
define the point in the fiber bundle.

Further, solving the parallel transport equation which is defined on the fiber as
\begin{eqnarray}
\frac{dx_\alpha(v)}{dv} \, {\cal D}_\alpha \textbf{g}(x(v)) = 0\,,
\end{eqnarray}
one can find the solution in terms of the Wilson line:
\begin{eqnarray}
\label{gg}
\textbf{g}(x)=Pexp\Big\{ ig \int\limits_{\mathbb{P}(x_0,x)} d\omega\cdot A(\omega)\Big\} \textbf{g}(x_0) ,
\end{eqnarray}
where the points $x_0$ and $x$ are connected by the path $\mathbb{P}$.
The starting point $x_0$ is usually fixed, {\it i.e.} $x_0$ is independent on $x$. Here,
$\textbf{g}(x_0)$ is chosen to be equal to unity
\footnote{The discussion regarding the choice of $x_0$ can be found in \cite{ContourG}.}.
Note that the fixing of $\textbf{g}(x)$ ensures a unique choice of the element in the orbit.
Inserting (\ref{gg}) into (\ref{Ag}), one can see that the field $A^{\textbf{g}}_\mu(x)$ is completely
determined by the form of the path which connects the starting and final points. Moreover, using (\ref{Ag}) and
(\ref{gg}), one obtains the property:
\begin{eqnarray}
\label{propAg}
Pexp\Big\{ ig \int\limits_{x_0}^{x} d\omega\cdot A^{\textbf{g}}(\omega)\Big\} =
\textbf{g}^{-1}(x) Pexp\Big\{ ig \int\limits_{x_0}^{x} d\omega\cdot A(\omega)\Big\}\textbf{g}(x_0) .
\end{eqnarray}
\textbf{Inserting} Eqn. (\ref{gg}) into Eqn.(\ref{Ag}) (we remind that the point $x_0$ is fixed),
we arrive at
\begin{eqnarray}
\label{cg1}
A^{\textbf{g}}_\mu(x)=\int\limits_{\mathbb{P}(x_0,x)} dz_\alpha \frac{\partial z_\beta}{\partial x_\mu}\,
\textbf{g}^{-1}(z) \,G_{\alpha\beta}(z| A)\, \textbf{g}(z)=
\int\limits_{\mathbb{P}(x_0,x)} dz_\alpha \frac{\partial z_\beta}{\partial x_\mu}\,
G_{\alpha\beta}(z| A^{\textbf{g}}) ,
\end{eqnarray}
where
\begin{eqnarray}
\label{Gcontour}
G_{\alpha\beta}(z| A^{\textbf{g}})\equiv G_{\alpha\beta}^{\textbf{g}}(z)
=\textbf{g}^{-1}(z) \,G_{\alpha\beta}(z| A)\, \textbf{g}(z)
\end{eqnarray}
The contour gauge condition demands that $\textbf{g}(x)$ is equal to
unity for all $x$ belonging to the base, {\it i.e.}
\begin{eqnarray}
\label{cg2}
[x,\,x_0]\stackrel{def}{=}Pexp\Big\{ ig \int\limits_{x_0}^{x} d\omega\cdot A(\omega)\Big\} = 1 ,
\,\,\, \forall x \in \mathbb{R}^4 .
\end{eqnarray}
Therefore, within the contour gauge, the field $A^{\textbf{g}}_\mu$ (see, (\ref{cg1})) becomes
\begin{eqnarray}
\label{cg3}
A^{c.g.}_\mu(x)=
\int\limits_{\mathbb{P}(x_0,x)} dz_\alpha \frac{\partial z_\beta}{\partial x_\mu}\,
G_{\alpha\beta}(z| A) ,
\end{eqnarray}
{\it i.e.} the gluon field $A^{\textbf{g}}_\mu$ is a linear functional of the tensor $G_{\mu\nu}$.
Let us \textbf{briefly} comment a choice of the boundary conditions for gluons. As is pointed out in \cite{ContourG},
the starting point $x_0$ of the path ${\mathbb{P}(x_0,x)}$ (see, (\ref{cg3})) is not well defined by construction.
In turn, this leads to the presence of the so-called residual gauge transformation.
To avoid this ambiguity,
it is natural to fix the uncertainty in Eqn. (\ref{cg3}) by $A_{\mu}(x_0)=0$. In other words,
the starting point is fixed, and it
is independent from the destination of the path ${\mathbb{P}(x_0,x)}$.

It is easy to see that the representations (\ref{Ag1}) and (\ref{Ag2})
(or the representations (\ref{Phi1}) and (\ref{Phi2}))
can be derived from (\ref{cg3}) by fixing of the path $\mathbb{P}(x_0,x)$ as a straight line
connecting the point $x$ with $\mp\,\infty$, respectively.
In fact, it means that the representations (\ref{Phi1}) and (\ref{Phi2}) correspond to two different
contour gauges and give us two different representations for two different functions $B^V_{-}(x_1,x_2)$
and $B^V_{+}(x_1,x_2)$. These two representations
are associated with the final and initial state interactions (Eqns.(\ref{B-fsi}) and (\ref{B-isi})).

Moreover, in order to get a concrete representation for gluons within the axial gauge,
%which is, as shown, a particular case of the contour gauge,
we can explicitly parametrize the straight
line between the points $x$ and $\pm\infty$ along the ``minus" light-cone direction $n^-$:
\begin{eqnarray}
x_\alpha(s)\Big|_{x}^{\pm\infty}  = x_\alpha \pm
n_\alpha \,\lim_{\epsilon\to 0} \frac{1-e^{-\epsilon s}}{\epsilon} \Big|_{0}^{+\infty}\, .
\end{eqnarray}
Then, using Eqn. (\ref{cg3}), one gets
\begin{eqnarray}
\label{Axg}
A^{ax_{\pm}}_\mu(x)=\mp\, n_\alpha \int\limits_0^\infty ds \,
G_{\alpha\mu}(x\pm ns)\, e^{-\epsilon s}\Big|_{\epsilon\to 0}\, .
\end{eqnarray}
Based  on the contour gauge, the representations (\ref{Axg})
determine two different contour gauges: $A^{ax_{+}}_\mu(x)$ corresponds to the straight
line between $x$ and $+\infty$, and $A^{ax_{-}}_\mu(x)$ corresponds to the line between
$-\infty$ and $x$. Moreover, their projections on the light-cone vector $n^-$ are the same ones:
\begin{eqnarray}
\label{axg2}
n\cdot A^{ax_\pm}(x)=A^+=0\, .
\end{eqnarray}

Thus, we can conclude that the {\it r.h.s.} of (\ref{Phi1}) and (\ref{Phi2})
correspond to the different functions  $B^V_{\pm}(x_1,x_2)$:
\begin{eqnarray}
\label{B-plus}
&&B^V_{+}(x_1,x_2)=\delta(x_1-x_2)B^V_{A(-\infty)}(x_1)+
\frac{T(x_1,x_2)}{x_1-x_2 + i\epsilon}\,\quad \text{for the gauge}\, \quad [x,\,-\infty]=1\,,
\\
\label{B-minus}
&&B^V_{-}(x_1,x_2)=\delta(x_1-x_2)B^V_{A(+\infty)}(x_1)+
\frac{T(x_1,x_2)}{x_1-x_2 - i\epsilon}\, \quad \text{for the gauge} \,\quad [+\infty,\, x]=1\,.
\end{eqnarray}
It is clear that these two representations can not be equivalent of each other.
Also, it is reasonable to assume the zeroth boundary conditions for gluons to be
$B^V_{A(\pm\infty)}=0$
which is in agreement with Refs. \cite{ContourG, AT-10}. Note that the functions $B^V_{\pm}$ do not
possess a certain property under the
time-reversal transformation. This property becomes a well-defined one only after
calculation of the imaginary part
for the hadron tensor.

We thus have no the ambiguity in the solutions of (\ref{DiffEqnsG}). As a result,
the $B^V$-function has a non-trivial imaginary part which contributes to $H^{(b)}_{\mu\nu}$ needed for the
gauge-invariant set.

\section{Comparison with the perturbative Compton scatering amplitude}\label{App:B}

In this Appendix, we discuss an essential difference between the QCD gauge invariance for
the ``perturbative" Comptom amplitude, $\langle q(k)g(\ell)|\, \mathbb{S} \,| g(k_2) q(k_1)\rangle$,
and the ``nonperturbative" analog of Compton amplitude, $\langle X(P_X)q(k)|\, \mathbb{S} \,| g(k_2) A(P)\rangle$,
where one of gluons is included in the loop integration (see, Figs.\ref{Compton}, \ref{ComptonNP}).

\subsection{QCD gauge invariance of the perturbative Compton amplitude}

\begin{figure}[t]
\centerline{\includegraphics[width=0.5\textwidth]{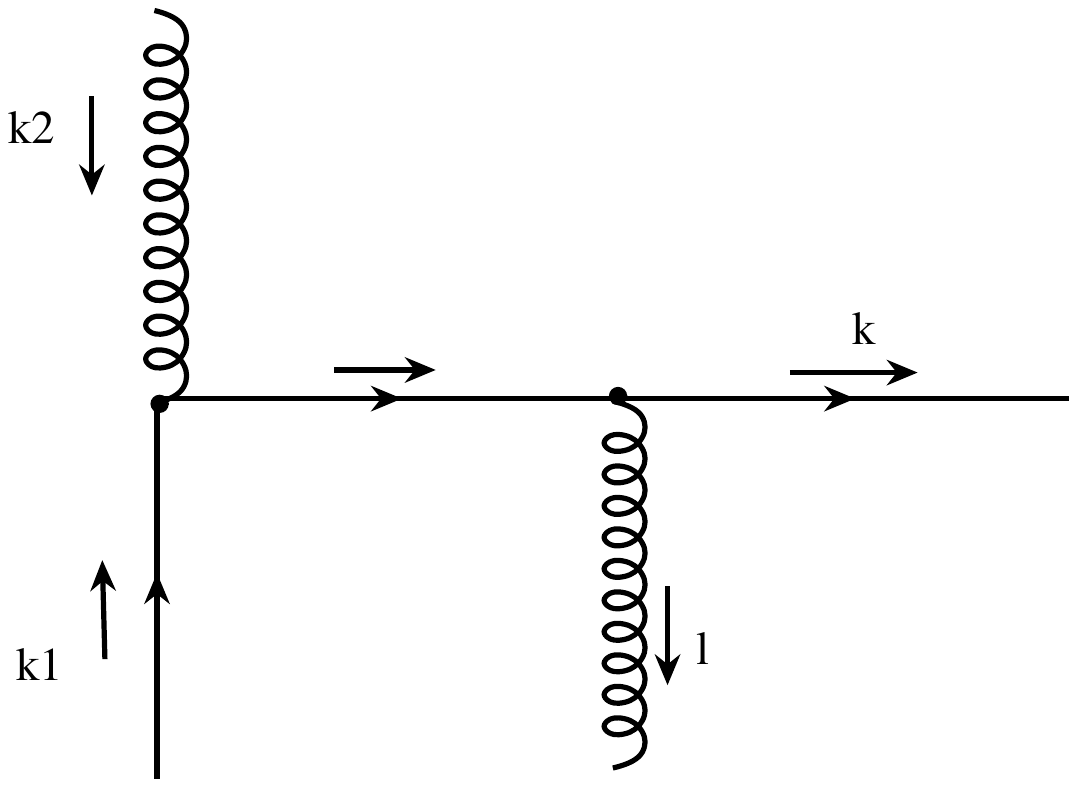}
\hspace{-3.cm}\includegraphics[width=0.5\textwidth]{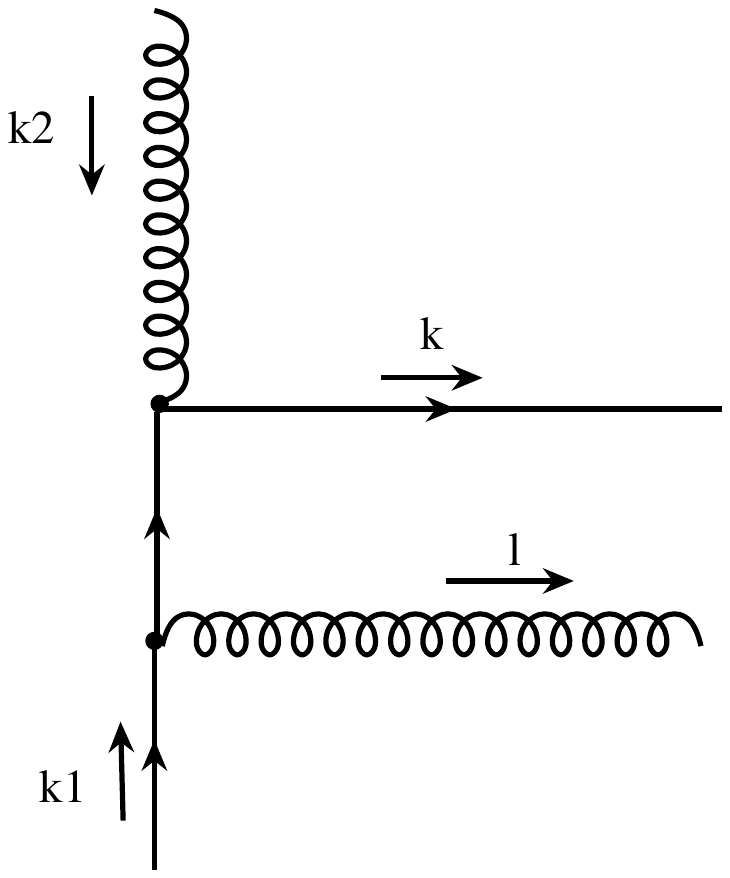}}
\vspace{-8.cm}
\centerline{\includegraphics[width=0.6\textwidth]{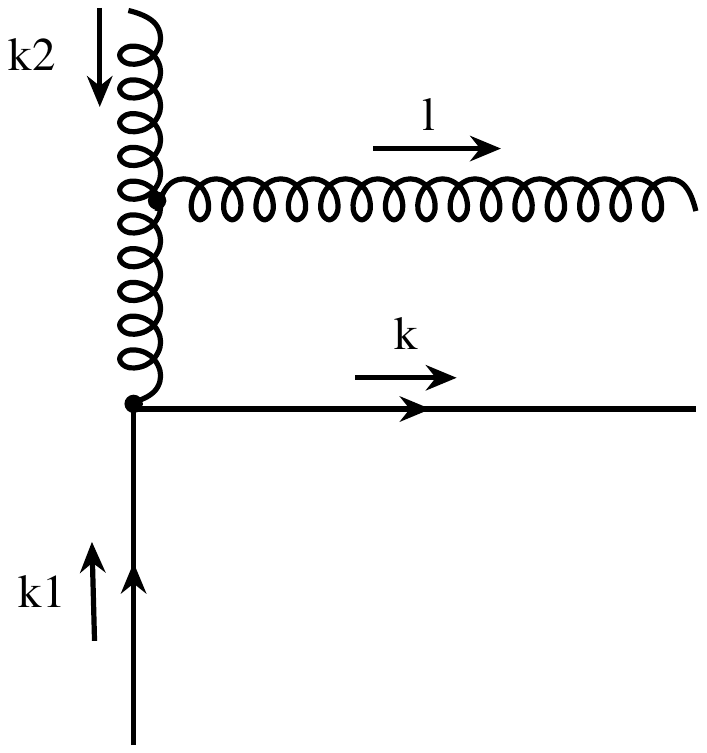}}
\vspace{-9.cm}
\caption{Compton diagrams without the blob (the ``perturbative" case).}
\label{Compton}
\end{figure}

We first consider the ``perturbative" Compton amplitude represented in Fig.\ref{Compton}.
%%%%%%%%%%%%%%%%%%%%%%%%%%%%% FIGURE %%%%%%%%%%%%%%%%%%%%%%%%%%%%%%%%
%%%%%%%%%%%%%%%%%%%%%%%%%%%%%%%%%%%%%%%%%%%%%%%%%%%%%%%%%%%%%%%%%%%%%%%
%
To proof the QCD gauge invariance, we assume that all gluons are physical ones with the transverse polarizations.
The first diagram in Fig.\ref{Compton} gives us
\begin{eqnarray}
\label{CompP-1}
{\cal A}_{\alpha\beta}(\text{dia.1})=(-i)g\,
\bar u(k) \gamma_\beta\,\frac{\hat k_1 + \hat k_2}{(k_1+k_2)^2+i\epsilon}\, \gamma_{\alpha} \,t^a\,t^b\, u(k_1)\,,
\quad {\cal A}(\text{dia.1})={\cal A}_{\alpha\beta}(\text{dia.1}) \epsilon^\perp_\alpha\, \epsilon^{*\,\perp}_\beta\,.
\end{eqnarray}
To check the gauge invariance with respect to, for example, the physical gluon with momentum $\ell$, we have to replace
the given polarization vector $\epsilon^{*\,\perp}_\beta$ on the momentum $\ell_\beta$, provided $\ell\cdot\epsilon^{*\,\perp}=0$,
in the amplitude (\ref{CompP-1}). We have
\begin{eqnarray}
\label{CompP-1-2}
{\cal A}(\text{dia.1})=(-i)g\,
\bar u(k) \hat\ell \, \frac{\hat k_1 + \hat k_2}{(k_1+k_2)^2+i\epsilon} \hat\epsilon^\perp \,t^a\,t^b\, u(k_1)\,.
\end{eqnarray}
Using the momentum conservation, $k_1+k_2=k+\ell$,  and the corresponding equations of motion,
the Eqn. (\ref{CompP-1-2}) reduces to the following form:
\begin{eqnarray}
\label{CompP-1-3}
{\cal A}(\text{dia.1})=(-i)g\,
\bar u(k)\, [\hat k_1 + \hat k_2 + \hat k] \,
\frac{\hat k_1 + \hat k_2}{(k_1+k_2)^2+i\epsilon} \hat\epsilon^\perp \,t^a\,t^b\, u(k_1)=
\bar u(k) \,\hat\epsilon^\perp \,t^a\,t^b\, u(k_1)\,.
\end{eqnarray}
In the similar way, we can get the contribution of the
second diagram presented in Fig.\ref{Compton}. It reads
\begin{eqnarray}
\label{CompP-1-33}
{\cal A}(\text{dia.2})=(-i)g\,
\bar u(k)\,\hat\epsilon^\perp \,
\frac{ \hat k_1 - \hat\ell }{ (k_1 - \ell)^2+i\epsilon} \hat\ell \,t^b\,t^a\, u(k_1)=
- \,\bar u(k) \,\hat\epsilon^\perp \,t^b\,t^a\, u(k_1)\,.
\end{eqnarray}
Therefore, the sum of these two diagrams gives us the color commutator:
\begin{eqnarray}
\label{CompP-12}
{\cal A}(\text{dia.1}) + {\cal A}(\text{dia.2})=
(-i)g\,\bar u(k) \,\hat\epsilon^\perp \,[t^a,\,t^b]\, u(k_1)\,.
\end{eqnarray}
In the Feynman gauge\footnote{For this discussion, the type of gauge is really not important.}
, the third diagram in Fig. \ref{Compton} contributes as (where we again replace
the transverse gluon polarization
on the gluon momentum: $\epsilon^{*\,\perp}\to\ell$)
\begin{eqnarray}
\label{CompP-3}
{\cal A}(\text{dia.3})=ig\, \bar u(k) \,\gamma_\sigma \,t^c\,  u(k_1) \frac{1}{(k-k_1)^2+i\epsilon}
if^{acb}V_{\alpha\sigma\beta}(-k_2; k-k_1; \ell) \epsilon^\perp_\alpha \ell_\beta\,,
\end{eqnarray}
where
\begin{eqnarray}
V_{\alpha\sigma\beta}(-k_2; k-k_1; \ell)= g_{\alpha\sigma} (k-k_1+k_2)_\beta +
g_{\sigma\beta}(\ell-k+k_1)_\alpha - g_{\alpha\beta}(k_2+\ell)_\sigma\,.
\end{eqnarray}
Making use of $\epsilon^\perp\cdot\ell=\epsilon^{*\,\perp}\cdot\ell=0$ and
\begin{eqnarray}
\frac{(k-k_1+k_2)\cdot \ell}{(k-k_1)^2}=\frac{k_2\cdot\ell}{(k_2-\ell)^2}=-1\,,
\end{eqnarray}
we obtain
\begin{eqnarray}
\label{CompP-3-2}
{\cal A}(\text{dia.3})=i^2g\, f^{abc} \bar u(k)\hat\epsilon^\perp\, t^c\, u(k_1)\,.
\end{eqnarray}
The sum of all three diagrams leads to
\begin{eqnarray}
\label{QCD-inv-Comp-P}
{\cal A}(\text{dia.1})+{\cal A}(\text{dia.2})+{\cal A}(\text{dia.3})=0\,
\end{eqnarray}
which ensures the QCD gauge invariance. In the case of the ``perturbative" Compton amplitude with
physical gluons, the momentum conservation, $\delta^{(4)}(k_1+k_2-\ell-k)$, and
the on-shellness of all external particles play important role for checking of the gauge invariance.

\subsection{QCD gauge invariance of the nonperturbative Compton scattering amplitude}

\begin{figure}[t]
\centerline{\includegraphics[width=0.5\textwidth]{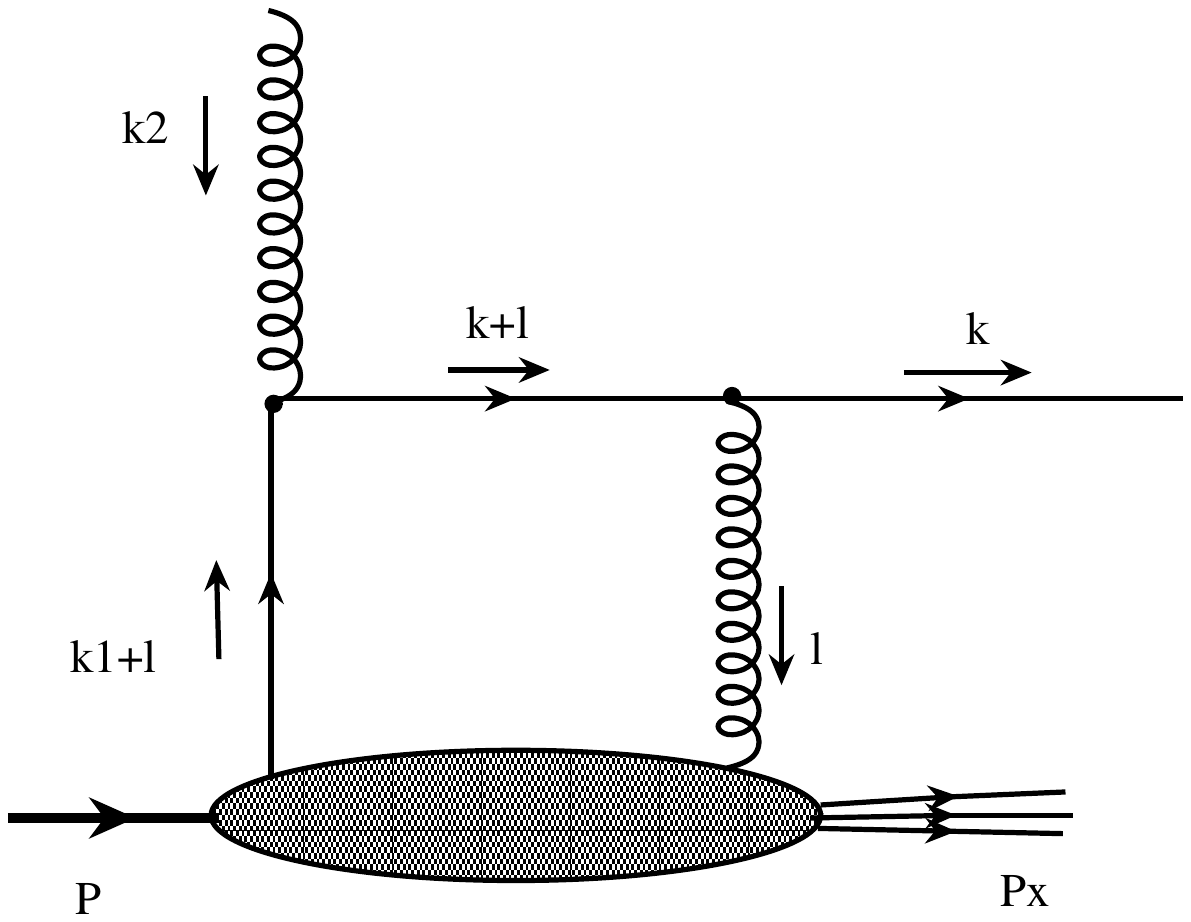}
\hspace{-2.cm}\includegraphics[width=0.5\textwidth]{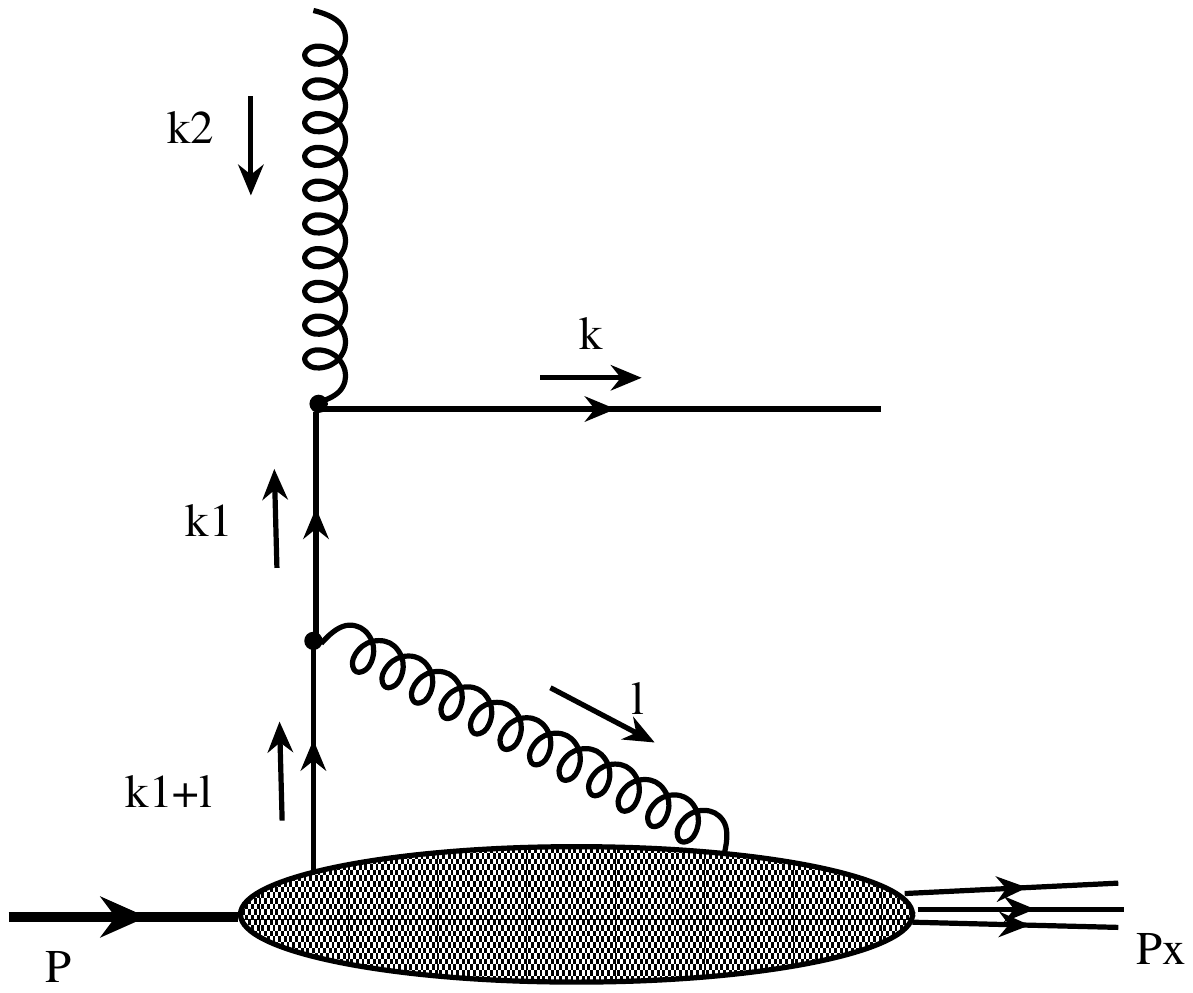}}
\vspace{-6.cm}
\caption{Compton diagrams with the blob (the ``nonperturbative" case).}
\label{ComptonNP}
\end{figure}

In order to get the diagrams
in Fig.~\ref{ComptonNP}, we now attach the nonperturbative blobs to the diagrams in Fig. \ref{Compton}.

In contrast to the ``perturbative" Compton amplitude, the gluon with
momentum, $\ell$, with respect to which we check the gauge invariance, is included in the loop integration.
As a result, the momentum conservation for the subprocess has a form: $k_1+k_2=k$.

Let us focus on two diagrams in Fig. \ref{ComptonNP} which are enough
to demonstrate how the color commutator contribution can be formed. Both diagrams are given by
\begin{eqnarray}
\label{NP-Compton}
\langle X(P_X)q(k)|\, \mathbb{S} \,| g(k_2) A(P)\rangle = \bar u(k)\epsilon^\perp_\alpha
\int(d^4\xi)(d^4\eta) e^{-ik_2\xi+ik\eta} \langle P_X | \frac{\delta^2 \mathbb{S}}{\delta\bar\psi(\eta)\delta A_\alpha(\xi)}| P\rangle
\Big|_{\psi=...=A=0}\,.
\end{eqnarray}
The first diagram in Fig. \ref{ComptonNP} contributes as
\begin{eqnarray}
\label{NP-Comp-1}
{\cal B}(\text{dia.1})= \bar u(k) \int(d^4\ell) \gamma_\beta S(\ell+k)\hat\epsilon^\perp
\int(d^4\eta) e^{-i\ell\eta} \langle P_X| A_{\beta}(\eta) \,t^a\,t^b\, \psi(0)|P\rangle\,.
\end{eqnarray}
This expression corresponds to the amplitude before factorization and, therefore,
the quark with $k_1+\ell$ and gluon with $\ell$ are off-shell. Next, we apply the factorization procedure and derive the following expression
for the first diagram contribution (here we use the light-cone basis presented above):
\begin{eqnarray}
\label{NP-Comp-1-2}
&&{\cal B}(\text{dia.1})=
\\
&&\bar u(k) \int dx_2 (x_{2}-x_1)\hat n^* S\big((x_2-x_1)n^* + k\big)\hat\epsilon^\perp\, t^a\,t^b\,
u(x_2n^*) \langle P_X | a^+\big((x_2-x_1)n^*\big) b^-(x_2n^*) |P\rangle
\nonumber
\end{eqnarray}
or
\begin{eqnarray}
\label{NP-Comp-1-2-2}
{\cal B}(\text{dia.1})=\bar u(k)\int dx_2 \gamma^-\,\gamma^+ \hat\epsilon^\perp \, t^a\,t^b\, u(x_2n^*)
\langle P_X | a^+\big((x_2-x_1)n^*\big) b^-(x_2n^*) |P\rangle\,,
\end{eqnarray}
where we replace the transverse gluon polarization vector by the gluon momentum $(x_2-x_1)n^*$.
In Eqn.~(\ref{NP-Comp-1-2-2}), we use the Fourier transformations for the quark and gluon fields.
$a^+$ and $b^-$ denote the gluon creation and
quark annihilation operators, respectively.

In the similar manner, we consider the second diagram contribution of Fig.~\ref{ComptonNP}.
Before factorization, we have
\begin{eqnarray}
\label{NP-Comp-2}
{\cal B}(\text{dia.2})=
\bar u(k)\hat\epsilon^\perp S(k_1) \int(d^4\ell) \hat\ell \,t^b\,t^a\, u(k_1+\ell)
\langle P_X| a^+(\ell) b^-(k_1+\ell)|P\rangle\,.
\end{eqnarray}
Having applied the factorization procedure, we derive the following expression:
\begin{eqnarray}
\label{NP-Comp-2-2}
{\cal B}(\text{dia.2})=
\bar u(k)\int dx_2\hat\epsilon^\perp  \gamma^-\,\gamma^+ \,t^b\,t^a\, u(x_2n^*)
\langle P_X| a^+\big((x_2-x_1)n^*\big) b^-(x_1n^*) |P\rangle\,.
\end{eqnarray}
Analyzing Eqns. (\ref{NP-Comp-1-2-2}) and (\ref{NP-Comp-2-2}), one can see that
in order to form the color commutator combination we have to insist on the ``external" conditions which actually emanate
from the presence of gluon poles. Indeed, if the gluon poles are present, the amplitude ${\cal B}(\text{dia.1})$ is accompanied
by the factor of $i\pi\delta(x_2-x_1)$. At the same time,
the amplitude ${\cal B}(\text{dia.2})$ has the factor of $-i\pi\delta(x_2-x_1)$
in according to the FSI and ISI prescriptions, see (\ref{QCD-illust}).

Thus, to check the QCD gauge invariance, the case of the ``perturbative" Compton amplitude with the physical gluons in the
initial and final states does not need the external condition for the presence of gluon poles. It is sufficient to use only
the momentum conservation, $k_1+k_2=\ell+k$, and the equations of motion for the initial and final quarks.
At the same time, in order to demonstrate the QCD gauge invariance for the case represented in Fig.~\ref{ComptonNP}
where one of gluon momenta belongs to the loop integration, the existence of the gluon poles with the corresponding FSI and ISI
prescriptions must be included.

%%%%%%%%%%%%%%%%%%%%%%%%%%%%% FIGURE %%%%%%%%%%%%%%%%%%%%%%%%%%%%%%%%
%%%%%%%%%%%%%%%%%%%%%%%%%%%%%%%%%%%%%%%%%%%%%%%%%%%%%%%%%%%%%%%%%%%%%%%

%%%%%%%%%%%%%%%%%%%%%%%%%%%%%%%%%%%%%%%%%%%%%%%%%%%%%%%%%%%%%%%%%%%%%%%%%%%%

\end{document}